\documentclass[twocolumn,a4paper,10pt]{article}
\usepackage{amsmath}
\usepackage{amsfonts}
\usepackage{amssymb}
\usepackage{latexsym}
\usepackage{epsfig}
\usepackage{graphicx}
\usepackage{oldgerm}
\usepackage{theorem}
\usepackage{bm}

\usepackage[hang,small,bf]{caption}
\usepackage[subrefformat=parens]{subcaption}
\def\N{\mathbb{N}}
\def\C{\mathbb{C}}
\def\R{\mathbb{R}}
\def\T{\mathbb{T}}
\def\D{\mathbb{D}}

\def\bra{\langle}
\def\ket{\rangle}

\def\x{\bm{x}}
\def\bLambda{\bm{\Lambda}}
\def\1{{\bf 1}}

\def\u{\bm{u}}
\def\v{\bm{v}}

\def\cD{\mathcal{D}}

\def\rO{{\rm O}}
\def\rN{{\rm N}}

\def\rR{{\rm R}}
\def\rI{{\rm I}}

\theorembodyfont{\itshape}

\newtheorem{thm}{Theorem}[section]
\newtheorem{lem}[thm]{Lemma}

\newtheorem{prop}[thm]{Proposition}

\newtheorem{con}[thm]{Conjecture}

\newcommand{\SSC}[1]{\section{#1}\setcounter{equation}{0}}
\newcommand{\qed}{\hbox{\rule[-2pt]{3pt}{6pt}}}

\newfont{\bg}{cmr10 scaled\magstep2}
\newcommand{\bigzerol}{\smash{\hbox{\bg 0}}}
\newcommand{\bigzerou}{\smash{\lower1.7ex\hbox{\bg 0}}}

\begin{document}

\title{\bf 
Eigenvalue and pseudospectrum processes \\
generated by nonnormal Toeplitz matrices \\
with rank 1 perturbations
}

\author{
Saori Morimoto$^{\dagger}$, \, 
Makoto Katori$^{\dagger,\S}$ \
and 
Tomoyuki Shirai$^{\ddagger,\P}$
\\
\\
$^{\dagger}$Department of Physics,
Faculty of Science and Engineering, \\
Chuo University,
Kasuga, Bunkyo-ku, Tokyo 112-8551, Japan \\
$^{\ddagger}$Institute of Mathematics for Industry, 
Kyushu University, \\
744 Motooka, Nishi-ku,
Fukuoka 819-0395, Japan \\
\\
$^{\S}$makoto.katori.mathphys@gmail.com, 
\,
$^{\P}$shirai@imi.kyushu-u.ac.jp
}

\date{6 December 2025}
\pagestyle{plain}

\twocolumn[

\maketitle

\vskip 1cm

\begin{abstract}
We introduce two kinds of matrix-valued dynamical processes
generated by nonnormal Toeplitz matrices with 
the additive rank 1 perturbations $\delta J$, 
where $\delta \in {\mathbb{C}}$ and $J$ is the all-ones matrix.
For each process, 
first we report the complicated motion
of the numerically obtained eigenvalues.
Then we derive the specific equation which determines
the motion of non-zero simple eigenvalues and clarifies the 
time-dependence of degeneracy 
of the zero-eigenvalue $\lambda_0=0$.
Comparison with the solutions of this equation,
it is concluded that the numerically observed non-zero eigenvalues
distributing around $\lambda_0$ are 
the exact eigenvalues not of the original system, 
but of the system perturbed by 
uncontrolled rounding errors of computer.
The complex domain in which the eigenvalues of randomly
perturbed system are distributed is identified with
the pseudospectrum including $\lambda_0$
of the original system with $\delta J$.
We characterize the pseudospectrum processes
using the symbol curves of the corresponding
nonnormal Toeplitz operators without $\delta J$. 
We report new phenomena
in our second model such that at each time
the outermost closed simple curve 
cut out from the symbol curve
is realized as the exact eigenvalues,
but the inner part of symbol curve
is reduced in size and embedded in the pseudospectrum
including $\lambda_0$.
Such separation of exact simple eigenvalues and 
a degenerated eigenvalue associated with pseudospectrum
will be meaningful for numerical analysis,
since the former is stable and robust, 
but the latter is highly sensitive and unstable
with respective to perturbations.
The present study will be related to the pseudospectra 
approaches to non-Hermitian systems developed in 
quantum physics

\vskip 0.5cm


\vskip 1cm

\end{abstract}
]

\vspace{3mm}

\clearpage


\SSC{Introduction}
\label{sec:introduction}

In this paper we will study the relationship between 
eigenvalues and pseudospectra,
which is an interesting and important topic
from the perspectives of numerical analysis \cite{CR01,HP05,TE05,WMG08}
and non-Hermitian quantum physics \cite{AGU20,GAKTHU18,OKSS20}. 
The present work was motivated by recent studies
on a time-dependent random matrix model
called the \textit{non-Hermitian 
matrix-valued Brownian motion} (BM) 
\cite{BCH24,BD20,Burda14,Burda15,EKY23,Fyo18,GW18}. 
Let $n \in \N:=\{1,2, \dots\}$. 
Consider $2n^2$ independent one-dimensional standard BMs,
$(B^{\rR}_{jk}(t))_{t \geq 0}$,
$(B^{\rI}_{jk}(t))_{t \geq 0}$, $1 \leq j, k \leq n$. 
Let $i:=\sqrt{-1}$ and we define the 
$n \times n$ non-Hermitian matrix-valued BM by
\begin{align*}
M(t) &=(M_{jk}(t))_{1 \leq j, k \leq n}
\nonumber\\
&:= \left( \frac{1}{\sqrt{2n}} (
B^{\rR}_{jk}(t)+ i B^{\rI}_{jk}(t)) \right)_{1 \leq j, k \leq n},
\, t \geq 0, 
\end{align*}
which starts from a deterministic matrix,  
$M(0)=(M_{jk}(0))_{1 \leq j, k \leq n} \in \C^{n^2}$.
We consider the \textit{eigenvalue process}
of $(M(t))_{t \geq 0}$ denoted by
$\bLambda(t)=(\Lambda_j(t))_{j=1}^n \in \C^n$,
$t \geq 0$. 
When the matrix-valued BM 
starts from the null matrix, $M(0)=O$,
$(\bLambda(t))_{t \geq 0}$
starts from the $n$ particles all degenerated at 
the origin, $n \delta_0$, and exhibits a uniform distribution
in an expanding disk centered at the origin 
with radius $\sqrt{t}$ on a complex plane $\C$,
$t > 0$ (\textit{the circular law}) \cite{Bai97,Gir84,TV08}.
At each time $t >0$, $\Lambda(t)$ is 
identified with the 
\textit{complex Ginibre ensemble} of eigenvalues
with variance $t$ \cite{Gin65}, which has been 
extensively studied 
in random matrix theory \cite{BF24,For10}.

\begin{figure}[htbp]
    \begin{tabular}{ccc}
    \hskip -0.3cm
      \begin{minipage}[t]{0.28\hsize}
       \centering
        \includegraphics[keepaspectratio, scale=0.15]{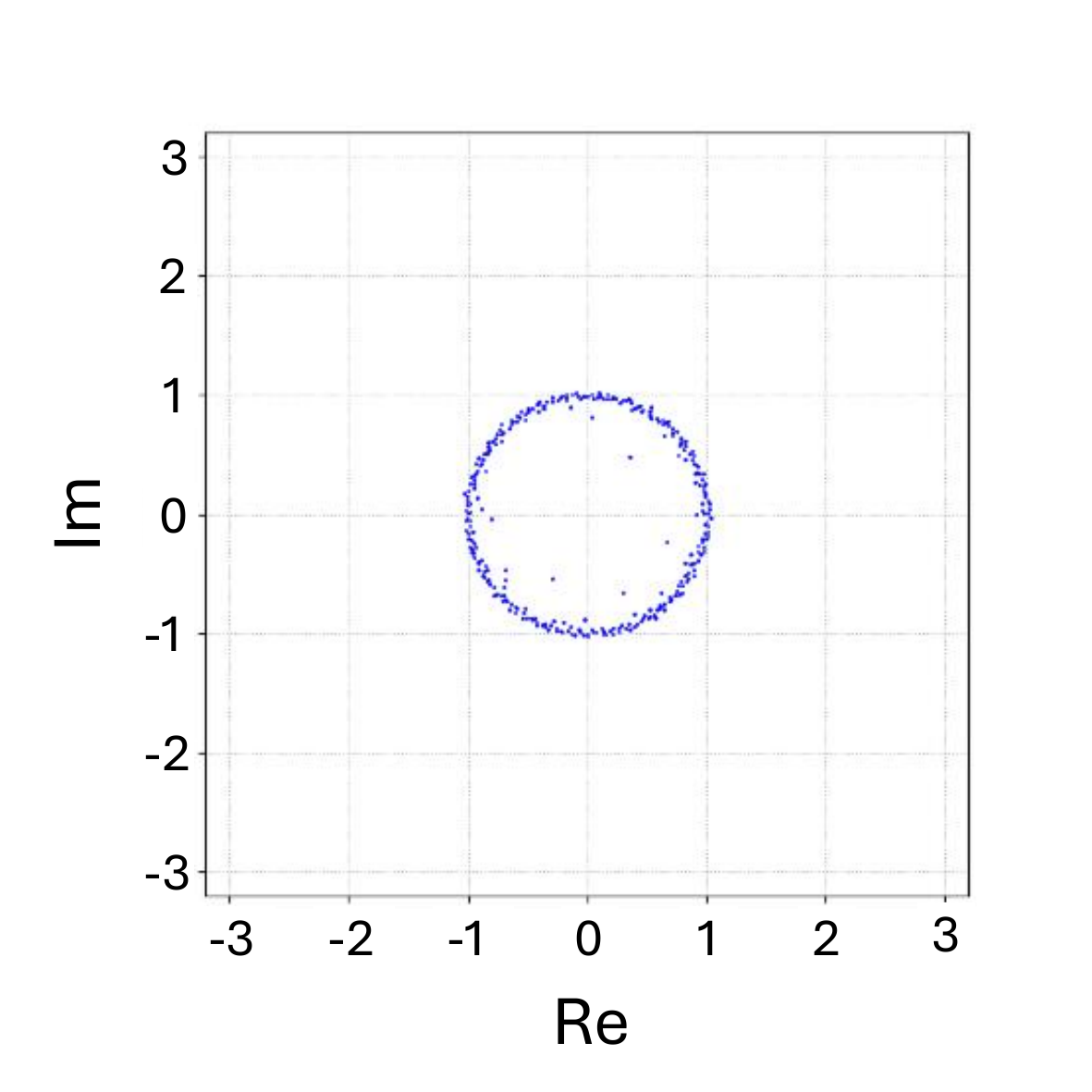}
       \subcaption{}
        \label{fig:ring}
      \end{minipage} 
      &
      \begin{minipage}[t]{0.28\hsize}
       \centering
        \includegraphics[keepaspectratio, scale=0.15]{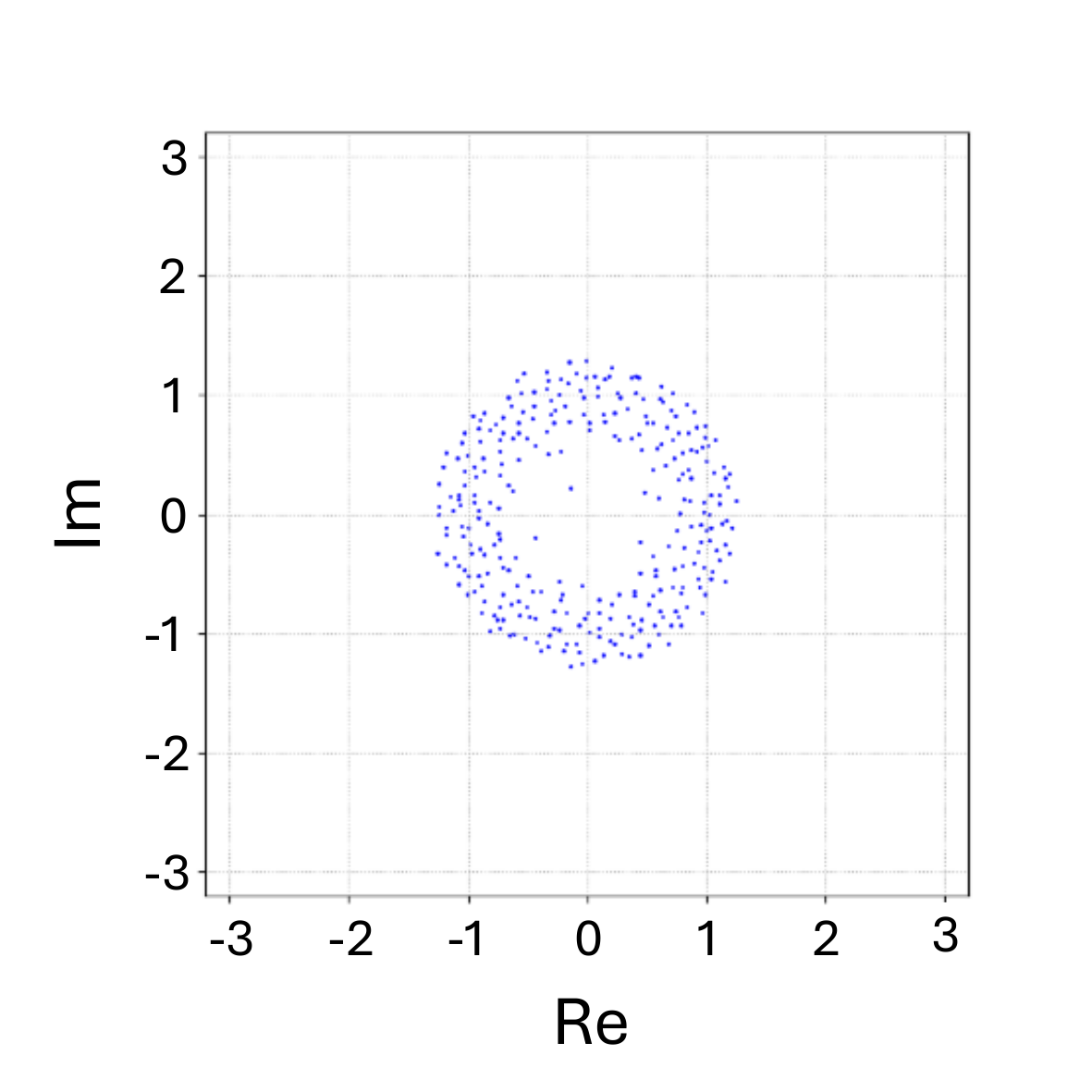}
        \subcaption{}
        \label{fig:annulus}
      \end{minipage} 
      &
      \begin{minipage}[t]{0.28\hsize}
       \centering
        \includegraphics[keepaspectratio, scale=0.15]{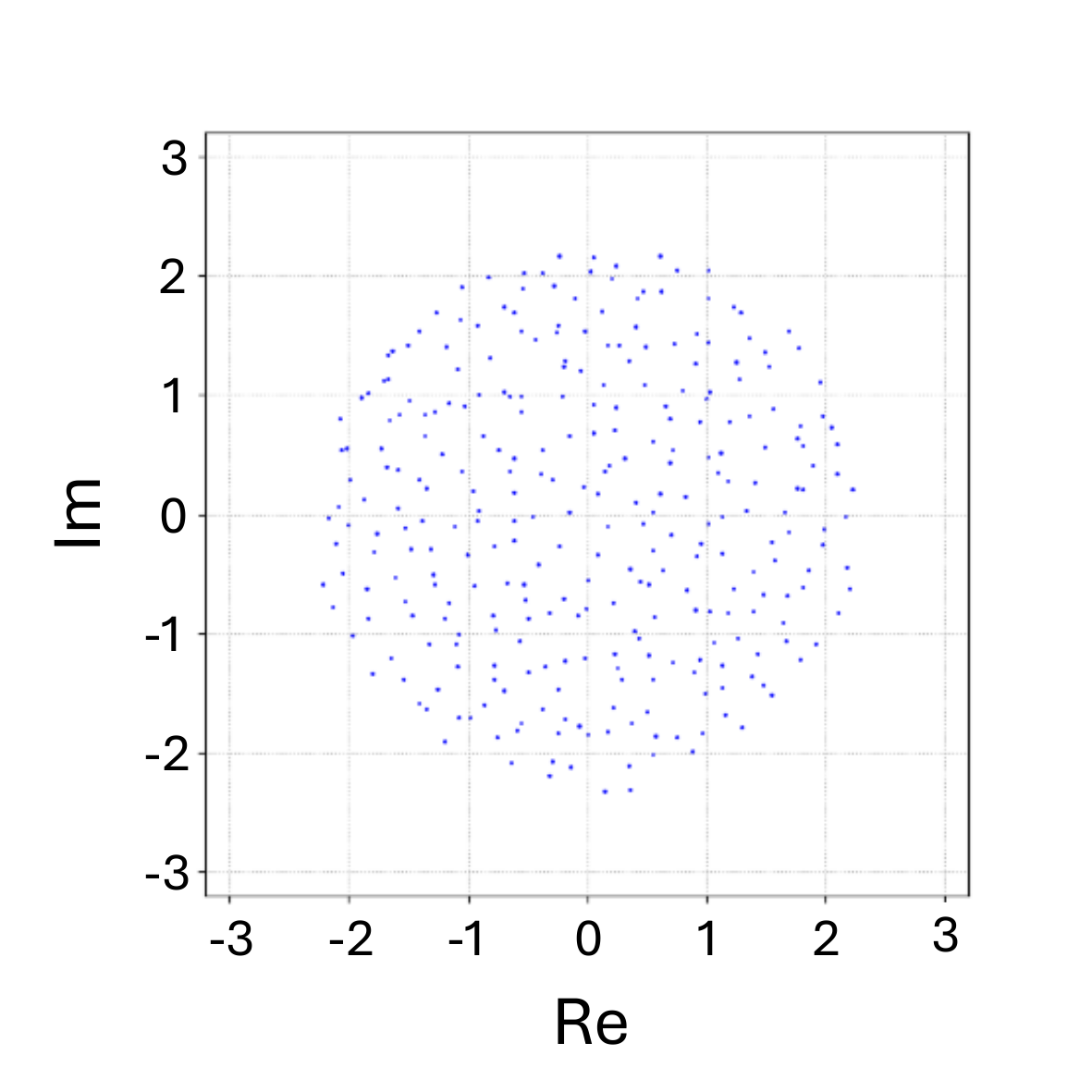}
        \subcaption{}
        \label{fig:disk}
      \end{minipage} 
    \end{tabular}
     \caption{Numerically obtained 
     eigenvalues are dotted 
     for the non-Hermitian 
     matrix-valued Brownian motion,  
     $(M(t))_{t \geq 0}$, with size $n=300$
     starting from the nonnormal matrix $S$.
     (a) Ring structure ($0<t \ll 1$), 
     (b) Growing annulus ($0 < t < 1$), 
     (c) Fulfilled disk ($t \geq 1$). }
\label{fig:Fig1}
\end{figure}

Burda et al.~\cite{Burda15} studied the process starting from
$M(0)=S$,
where $S$ denotes the 
$n \times  n$ \textit{shift matrix} 
\begin{equation}
S:= (\delta_{j \, k-1} )_{1 \leq j, k \leq n}
=
\begin{scriptsize}
\left(
\begin{array}{cccccc}
0 & 1 &  &  & \bigzerou & \\
   & 0  & 1 & & & \\
& \ldots & \ldots & & & \\
& & \ldots & \ldots &&  \\
& &  & 0 & 1 &  \\
& &  &   & 0 & 1 \\
& \bigzerol & & & & 0
\end{array}
\right).
\end{scriptsize}
\label{eq:S}
\end{equation}
Here $\delta_{j k}$ is the Kronecker delta.
Notice that this matrix is \textit{nonnormal}; 
$S^{\dagger} S \not= S S^{\dagger}$.
(In the present paper, for
a square matrix $A \in \C^{n^2}$,
$A^{\dagger}$ denotes the complex conjugate
of transpose of $A$;
$A^{\dagger}:=\overline{A^{\sf T}}$.)
All eigenvalues of $S$ are zero, 
and hence the initial state of
$(\bLambda(t))_{t \geq 0}$ is 
$n \delta_0$, which is the same as 
that in the case $M(0)=O$.
By numerical simulation, however, 
Burda et al.~\cite{Burda15} found that 
the eigenvalues seem to expand instantly from
$n \delta_0$ to make a unit circle as dotted 
in Fig.~\ref{fig:ring}.
For the time interval $0 < t < 1$,
the dots form the growing annulus
(Fig.~\ref{fig:annulus}).
Then the inner radius of the annulus shrinks to
zero at $t=1$ and dots fill up
a full disk (Fig.~\ref{fig:disk}).
The disk expands with radius $\sqrt{t}$, $t > 1$
and the dots tend to follow the circular law
in $t \gg 1$.

It was claimed that \cite{Burda15} the unit circle
observed at the early stage 
will not be composed of exact eigenvalues, 
but will represent a
\textit{pseudospectrum} \cite{BS99,RT92,TE05}.
For a complex-valued square matrix $A \in \C^{n^2}$
with an arbitrary but fixed $\varepsilon > 0$,
the $\varepsilon$-pseudospectrum of $A$ 
is defined as an open subset 
$\sigma_{\varepsilon}(A)$ of $z \in \C$ such that
\begin{equation}
\| (z I- A)^{-1} \| > \varepsilon^{-1}.
\label{eq:def1_PS}
\end{equation}
Here $I$ is the unit matrix of size $n$ and
the matrix $(z I-A)^{-1}$ is known as the 
\textit{resolvent} of $A$ at $z$. 
In the present paper, we assume that
$\| \cdot \|$ is given by the 2-norm,
$\| A \| :=
\max_{\x \in \C^n, \|\x\|_2=1} \| A \x\|_2$,
where $\| \x \|_2:=\sqrt{ \sum_{j=1}^n |x_j|^2}$
for vector $\x=(x_1, \dots, x_n) \in \C^n$.
If $\sigma(A)$ denotes the spectra 
(i.e., set of eigenvalues of $A$), 
then
$\| (z I -A)^{-1} \| = \infty$, 
$z \in \sigma(A)$.
Hence, by definition, the exact eigenvalue is contained
in the $\varepsilon$-pseudospectrum for every 
$\varepsilon > 0$. 
In other words, the eigenvalues of $A$ are recovered
from $\sigma_{\varepsilon}(A)$ as poles of
$(z-A)^{-1}$ in the $\varepsilon \to 0$ limit \cite{TE05}. 

It is proved that the above definition 
of $\varepsilon$-pseudospectra with \eqref{eq:def1_PS} 
is equivalent with
the following definition 
\cite{RT92} \cite[Theorem 2.1]{TE05}. 
The $\varepsilon$-pseudospectrum of a matrix
$A \in \C^{n^2}$ with $\varepsilon >0$ is the set
of $z \in \C$ such that 
$z \in \sigma(A+E)$
for some matrix $E \in \C^{n^2}$ with
$\|E\| < \varepsilon$.
That is, for a given matrix, 
the $\varepsilon$-pseudospectrum is \textit{not}
the exact spectrum of the original matrix $A$,
but it is the set of exact eigenvalues of 
some perturbed matrix $A+E$ with $\|E\| < \varepsilon$.
We notice that, at $0 < t \ll 1$, 
the matrix $M(t)$ starting
from \eqref{eq:S} will be well approximated by
\begin{equation}
S+ \sqrt{ \frac{t}{2n} } Z, 
\label{eq:S+Z}
\end{equation}
where $Z=(Z_{jk})_{1 \leq j, k \leq n}$ with
the elements given by independent
complex Gaussian random variables, 
\[
Z_{jk}=X_{jk}+i Y_{jk},
\, 
X_{jk} \sim \rN(0,1),
\, 
Y_{jk} \sim \rN(0,1),
\]
$1 \leq j, k \leq n$.
The equivalence of the two definitions of pseudospectrum
mentioned above suggests the following: 
The eigenvalues with the ring structure reported by
Burda et al.~\cite{Burda15}
will be the eigenvalues of the
randomly perturbed matrix \eqref{eq:S+Z}
and the domain surrounded by the ring
is the pseudospectrum including the
highly degenerated zero-eigenvalue
$\lambda_0=0$ of $M(0)=S$ \cite{MKS24b}.
As a generalization of the randomly
perturbed system \eqref{eq:S+Z}, 
we are interested in the
discrete-time random process, 
\begin{equation}
S_{\delta Z}(m)
:=S^m +\delta Z, \quad m=1, 2, \dots, n,
\label{eq:pZ}
\end{equation}
where the nonnormality
of the matrix is changing in time $m$. 

In the present paper, we introduce the
following two kinds of 
discrete-time \textit{dynamical systems}
generated by nilpotent Toeplitz matrices,
in which the random perturbations by $Z$ are replaced by
the \textit{deterministic rank 1 perturbations} \cite{For23},  
\begin{align*}
\mbox{\bf [model 1]} 
\quad S^{(1)}_{\delta J}(m)
&:=S^m +\delta J, 
\nonumber\\
\mbox{\bf [model 2]} 
\quad
S^{(2)}_{\delta J}(m) &= S^{(2)}_{\delta J}(m; a)
\nonumber\\
&:= S^m + a S^{m+1}+\delta J,
\end{align*}
$m=1, 2, \dots, n$. 
Here $\delta, a \in \C$ and 
$J=(J_{jk})_{1 \leq j, k \leq n}$ is the all-ones matrix;
$J_{jk} \equiv 1$, $1 \leq j, k \leq n$. 
The \textbf{model 1} is a deterministic 
version of \eqref{eq:pZ}
and \textbf{model 2} is 
its one-parameter ($a \in \C$) extension.
Notice that $S^{(\ell)}_{\delta J}(m)$, 
$\ell=1,2$ are
non-Hermitian;
$(S^{(\ell)}_{\delta J}(m))^{\dagger} \not= 
S^{(\ell)}_{\delta J}(m)$,
and \textit{nonnormal},
$(S^{(\ell)}_{\delta J}(m))^{\dagger} S^{(\ell)}_{\delta J}(m)
\not= 
S^{(\ell)}_{\delta J}(m) (S^{(\ell)}_{\delta J}(m))^{\dagger}$,
$m=1,2,\dots, n-1$. 
For nonnormal matrices, both of 
\textit{right-eigenvectors}
and \textit{left-eigenvectors} are needed to construct
eigenspaces associated with eigenvalues, 
and the \textit{overlap matrix} is defined using
the right- and left-eigenvectors. 
The overlap matrices play important roles in
a variety of fields in mathematics and physics
and have been extensively studied 
\cite{ATTZ20,BNST17,BF24,CM98,CEX24,FM02,FO22,FS12,FT21,JNNPZ99,MC00}.
In particular, 
in the recent study of non-Hermitian matrix-valued
stochastic processes, the analysis of 
the coupling between the eigenvalue processes
and the eigenvector-overlap processes is one of the
central topics 
\cite{BCH24,BD20,Burda14,Burda15,EKY23,Fyo18,GW18,Yab20}. 
The square roots of the diagonal elements
of the overlap matrix are especially called 
the \textit{condition numbers} of eigenvalues
and we notice that 
the pseudospectra can be evaluated using
the condition numbers 
(e.g.~the Bauer--Fike theorem \cite[Sects. 35 and 52]{TE05}). 
In the present paper, we study the relationship between 
the eigenvalue processes
and the pseudospectrum processes
for the two models.

For $m=2, 3, \dots, n$, both models have 
a degenerated eigenvalue $\lambda_0=0$. 
There are two notions of degeneracy of eigenvalues.
The \textit{algebraic multiplicity} is the number of
times the eigenvalue appears as a root of 
the characteristic polynomial of the matrix, while
the \textit{geometric multiplicity} is the dimension
of the linear space of the eigenvectors associated
to the eigenvalue.
If the algebraic multiplicity of an eigenvalue 
exceeds the geometric multiplicity,
then that eigenvalue 
is said to be \textit{defective} and the matrix
becomes \textit{nondiagoralizable} \cite{TE05}. 
In both models, 
we can prove that $\lambda_0$ is defective
if $m=2, 3, \dots, n-2$.
(See Remark 1 below and \cite{MKS24b}.)
The \textit{defectivity} of $\lambda_0$ 
is expressed by the size of 
$\varepsilon$-pseudospectrum including $\lambda_0$
for given $0< \varepsilon \ll 1$. 

If $\delta=0$, that is, the 
rank 1 perturbation $\delta J$ is not applied,
then the matrices of our models are reduced to
\textit{banded Toeplitz matrices} for any $n \in \N$
and their $n \to \infty$ limits define 
the \textit{banded Toeplitz operators}. 
For any banded Toeplitz operator, 
a closed curve called the 
\textit{symbol curve} is defined on the complex plane $\C$
and the $\varepsilon$-pseudospectra 
of the nonnormal banded Toeplitz matrices
are specified by the whole regions 
surrounded by the symbol curves.
(See Theorem \ref{thm:Toeplitz}
and Corollary \ref{thm:pseudoS}
below \cite{RT92,TE05}).
In the present paper 
we report new phenomena
in \textbf{model 2} with $\delta \not=0$ 
such that at each time 
$m=2,3, \dots n-2$, 
the outermost closed simple curve 
cut out from the symbol curve
is realized as the \textit{exact eigenvalues},
but the inner part of symbol curve
composed of several closed simple curves
osculating each other 
is \textit{reduced in size} and embedded as
a complicated structure in the \textit{pseudospectrum} 
including $\lambda_0$.
Such phenomena have not been reported 
in the previous mathematical study of 
banded Toeplitz matrices 
with random 
perturbations \cite{BGKS22,BKMS21,BPZ19,BPZ20,BZ20,SV21}.
The size reduction of the pseudospectrum
including $\lambda_0$ expresses
the \textit{relaxation process}
of defectivity of $\lambda_0$.
Our dynamical systems show
the transitions from \textit{far-from-normal} matrices
to \textit{near-normal} matrices.
Algebraic descriptions of such relaxation processes
of defectivity is studied in \cite{MKS24b}
by calculating the Jordan block decompositions
of the resolvents of the matrices at $\lambda_0$.
We expect that the present model study of 
time-evolutionary pseudospectra will give
a new point of view to the 
\textit{pseudospectra analysis} and related
methods in numerical analysis \cite{CR01,HP05,TE05,WMG08}. 
It is also expected that the present mathematical study 
will provide useful models for physical phenomena
studied in \textit{non-Hermitian quantum physics} 
\cite{AGU20,GAKTHU18,OKSS20}. 

The paper is organized as follows.
In Section \ref{sec:numerical} we report
the numerical observations for the
eigenvalue processes for the models.
In Section \ref{sec:ev} we give the
equations which determine the exact eigenvalues
in Theorem \ref{thm:ev1} for \textbf{model 1}
and in Theorem \ref{thm:evG1} for \textbf{model 2},
respectively. 
The properties of the solutions of these equations
are given by Propositions
\ref{thm:Rouche}, \ref{thm:ev2}, and \ref{thm:evG2}.  
The notions of symbols and symbol curves for
banded Toeplitz operators are introduced in
Section \ref{sec:pseudo}
and our numerical results are studied 
in referring to the theory of pseudospectra
for banded Toeplitz matrices.
There we study new phenomena in
pseudospectrum processes exhibiting 
\textit{separation} of symbol curves and 
\textit{dilatation} of their inner parts.
Section \ref{sec:asymptotics} is devoted to
reporting the asymptotics of the exact-eigenvalue processes
and pseudospectrum processes in infinite-matrix 
limit $n \to \infty$ 
(Propositions \ref{thm:model1_asym} 
and \ref{thm:model2_asym}). 
Concluding remarks and future problems are
given in Section \ref{sec:future}. 

\SSC{Numerical Observations of Processes}
\label{sec:numerical}
\subsection{Model 1}
\label{sec:model1_numerical}

We have performed numerical calculation of 
the eigenvalue processes of \textbf{model 1} 
with a given size of matrix $n \in \N$.
The obtained eigenvalues are plotted 
on $\C$ for each time $m =1,2, \dots, n$.

The observations are explained 
using the case with $n=200$ and $\delta=0.01$ below.

\begin{figure}[htbp]
\centerline{
\includegraphics[width=8.5cm]{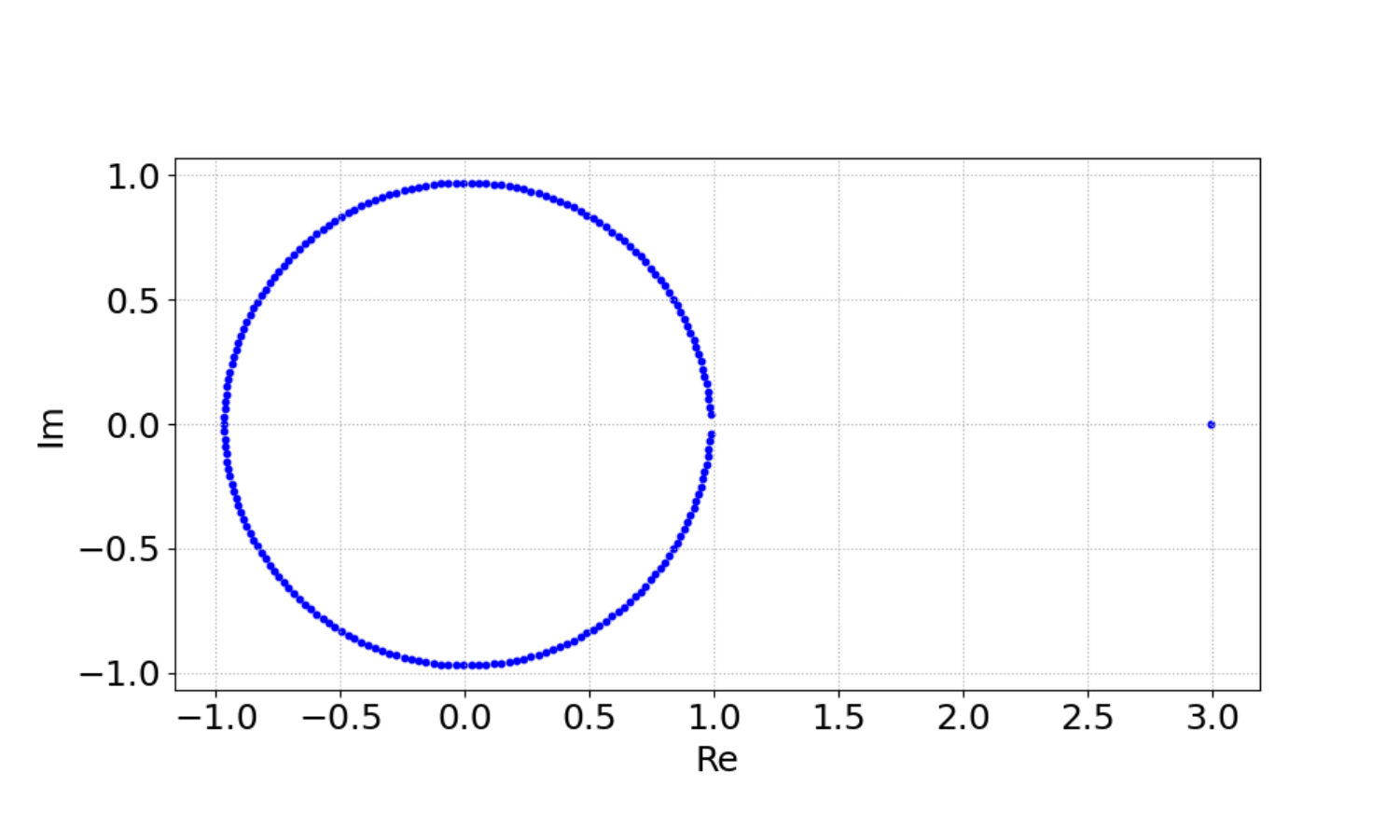}}
\caption{Plots of numerically obtained eigenvalues
at time $m=1$ 
for \textbf{model 1}, 
$S^{(1)}_{\delta J}(1)$, with
$n=200$ and $\delta=0.01$. 
199 dots form a unit circle missing one point at $z=1$, 
and an outlier is observed near $z=3$,
which is denoted by $\lambda_1(1)$.}
\label{fig:model1_numerical_m1}
\end{figure}

\begin{description}
\item{(i)} \quad
The numerically obtained eigenvalues are dotted in
Fig.~\ref{fig:model1_numerical_m1} at $m=1$. 
We find 199 dots which form
a unit circle missing one point at $z=1$, 
and one outlier located near $z=3$.
\end{description}

\noindent
As proved by Propositions \ref{thm:Rouche} (iia)
and \ref{thm:ev2} in Section \ref{sec:ev}, 
the dot near $z=3$ is identified with an exact eigenvalue
of $S_{\delta J}^{(1)}(1)$ 
and its time evolution, denoted by $(\lambda_1(m))_{m=1}^n$,
can be explicitly described using the Catalan numbers \cite{Rio68}, 
\begin{equation}
C_k := \frac{1}{k+1} \binom{2k}{k}
=\frac{(2k)!}{(k+1)! k!}, 
\label{eq:Catalan}
\end{equation}
$k \in \N_0:=\{0, 1, 2, \dots\}$, 
as
\begin{align}
\lambda_1(m) &=n \delta+ 1
\nonumber\\
& \quad -
\sum_{k=0}^{p_1-1} C_k 
\frac{(m/n)^{k+1}}{(n \delta)^k}
+\rO((n \delta)^{-p_1}),
\label{eq:outlierA}
\end{align}
for $1 \leq m \leq n-1$, 
where $p_1:=[(n-1)/m]$ 
(the greatest integer 
less than or equal to $(n-1)/m$), 
and
$\lambda_1(n)=n \delta$ (Proposition \ref{thm:ev2}).
So we will show mainly the eigenvalues which are
distinct from $\lambda_1(m)$ in the following figures.
(For outliers in spectra discussed in random matrix theory,
see \cite{For23,Tao13}.)

\begin{figure}[htbp]
    \begin{tabular}{cc}
      \begin{minipage}[t]{0.45\hsize}
        \centering
        \includegraphics[keepaspectratio, scale=0.20]
        {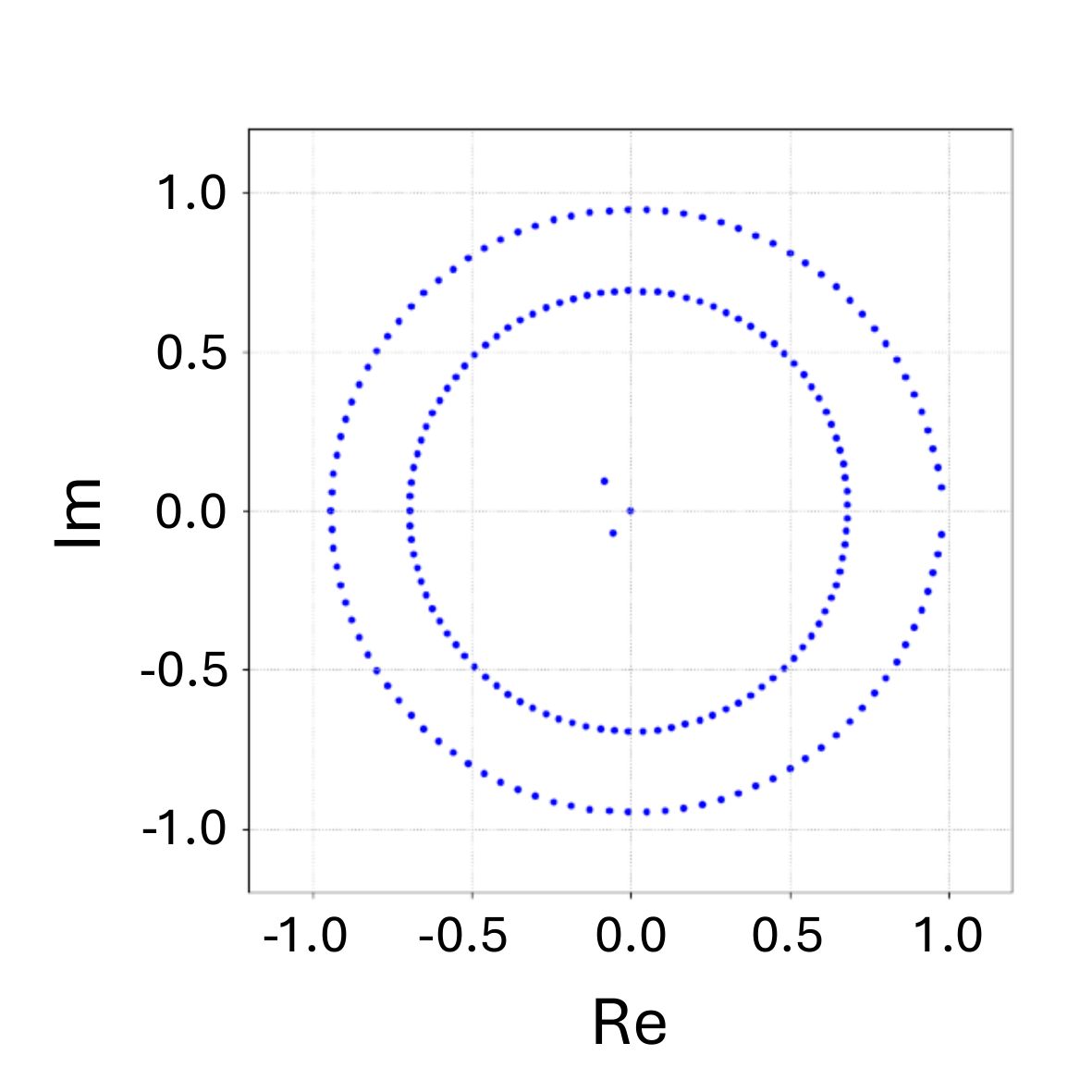}
        \subcaption{$m=2$}
        \label{fig:model1_numerical_m2}
      \end{minipage} &
      \begin{minipage}[t]{0.45\hsize}
        \centering
        \includegraphics[keepaspectratio, scale=0.20]
        {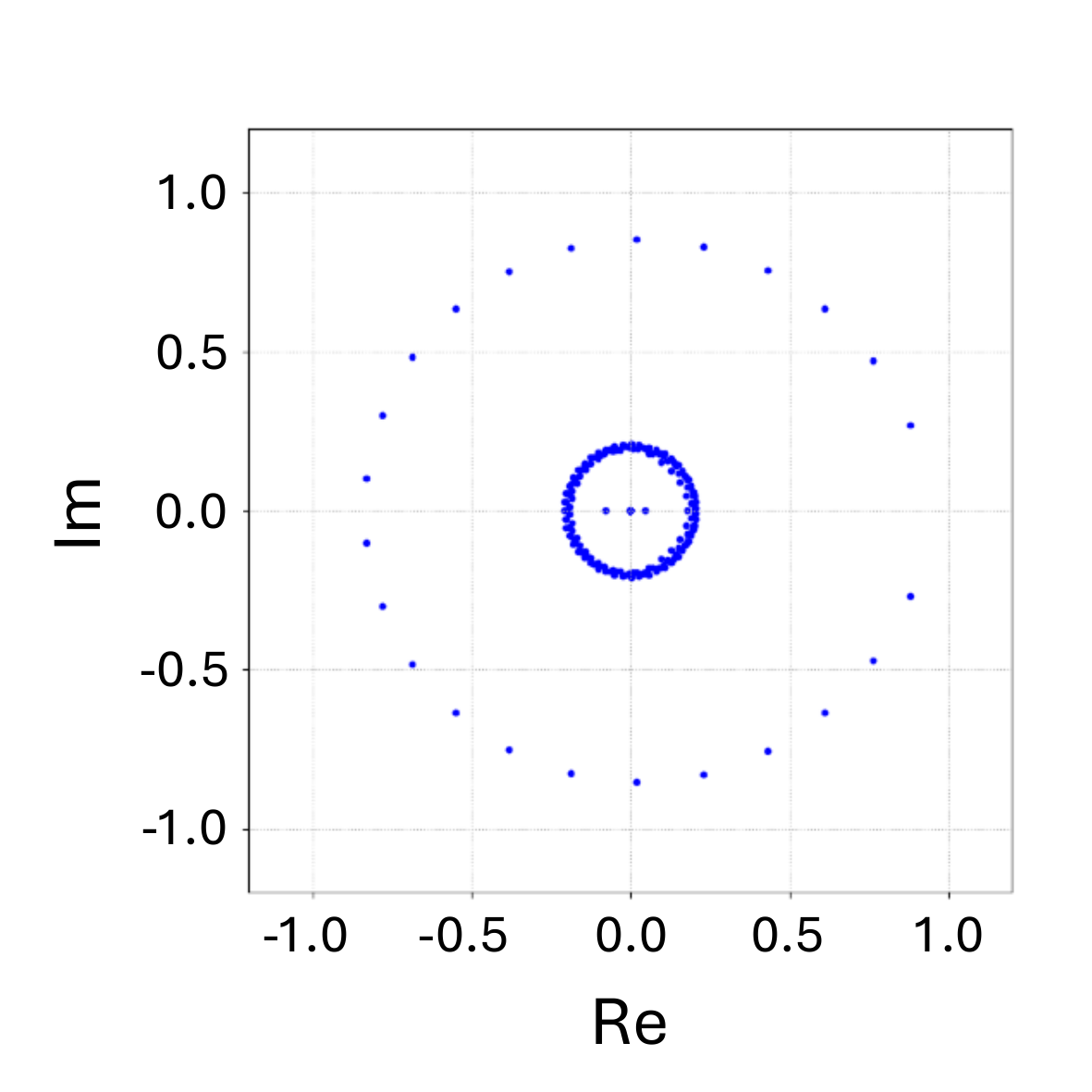}
        \subcaption{$m=8$}
        \label{fig:model1_numerical_m8}
      \end{minipage} \\
      \begin{minipage}[t]{0.45\hsize}
        \centering
        \includegraphics[keepaspectratio, scale=0.20]
        {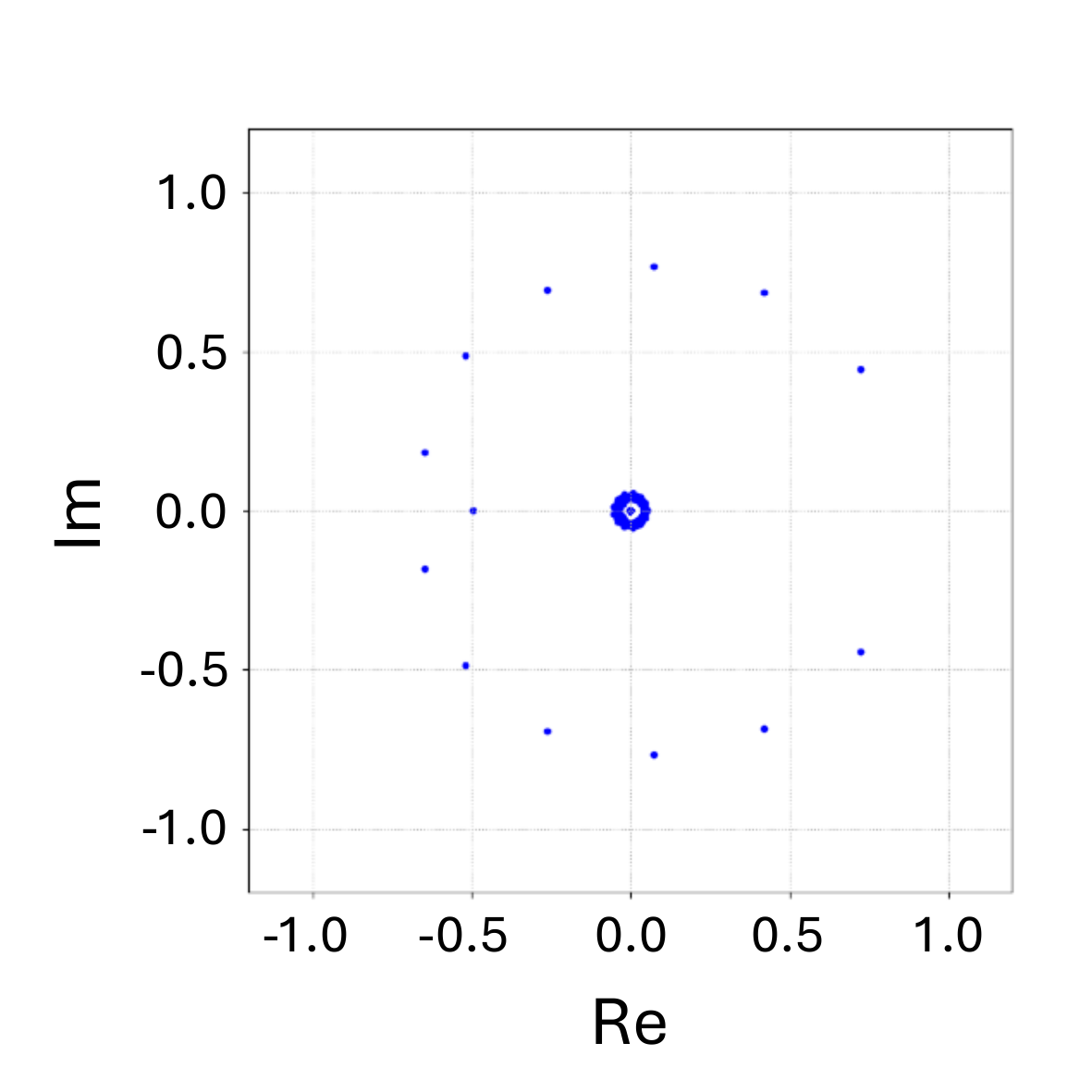}
        \subcaption{$m=15$}
        \label{fig:model1_numerical_m15}
      \end{minipage} &
      \begin{minipage}[t]{0.45\hsize}
        \centering
        \includegraphics[keepaspectratio, scale=0.20]
        {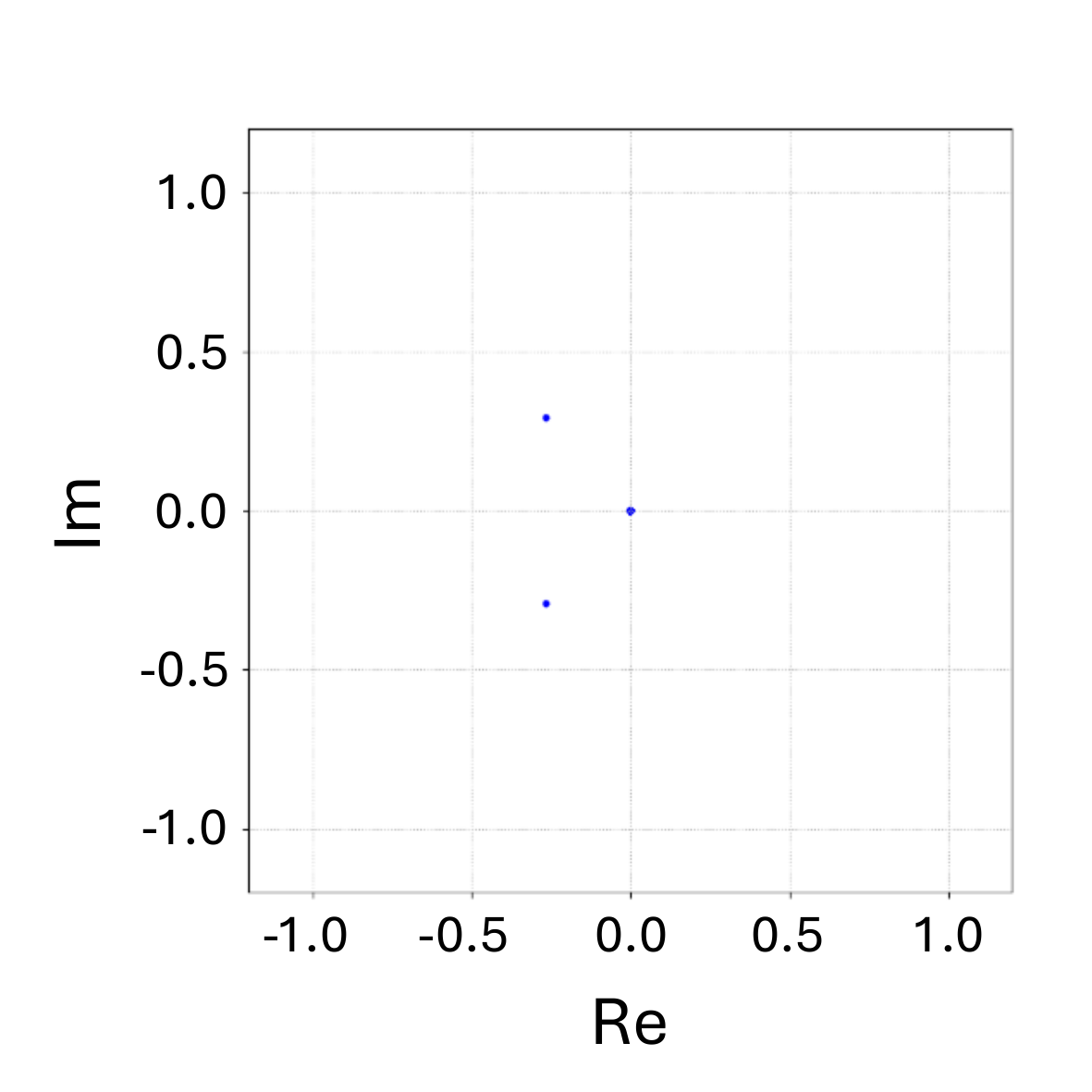}
        \subcaption{$m=80$}
        \label{fig:model1_numerical_m80}
      \end{minipage} 
    \end{tabular}
     \caption{Numerically obtained eigenvalues are
     plotted for \textbf{model 1}, 
     $(S^{(1)}_{\delta J}(m))_{m=1}^n$, 
     with $n=200$ and $\delta=0.01$ 
     at $m=2$, 8, 15, and 80, respectively.}
\label{fig:model1_numerical_ms}
  \end{figure}

Figure~\ref{fig:model1_numerical_ms} shows the 
numerical results for $m \geq 2$.

\begin{description}
\item{(ii)} \,
At time $m=2$, 99 dots form
a slightly deformed circle whose radius is $\lesssim 1$.
In addition to them
many dots appear and form a smaller circle
with radius $\simeq 0.7$.
In the vicinity of the origin, three dots are observed, 
one of which is located exactly at the origin.

\item{(iii)} \,
At time $m=8$, 24 dots form
an incomplete circle shaped `C', whose radius is about 0.8.
In addition, 
we see a small annulus whose boundaries are wavy. 
Within that smaller circle, we see three dots,
one of which is located at the origin. 

\item{(iv)} \,
As $m$ increases, the inner annulus shrinks to the origin.
The reduction of size is exponential as a function of $m$.
The outer dots in the upper (resp. lower) half plane of $\C$ 
move along a circle counterclockwise
(resp. clockwise) until they attach the negative real axis, 
$\R_- :=\{x \in \R; x < 0\}$.
Then they move along $\R_-$
repulsively with each other 
preserving the order of the distances
from the origin.
They are absorbed by the origin one by one.
At time $m=15$, only 13 dots remain apart from
the origin. One of them is on $\R_-$,
which will approach the origin 
and will be absorbed earlier than other 12 dots.

\item{(v)} \,
At time $m=80$, only two dots remain
apart from the dot at the origin.
In $m > 80$, both of the two dots approach
to $\R_-$, and then
they show repulsive motion on $\R_-$.
One of them is absorbed by the origin at $m=100$.

\item{(vi)} \,
At the final time $m=n=200$, there are only two dots, one of
them is at the origin, and the other one
is $\lambda_1(n)$ located at $z=2$.
They are the eigenvalues of $\delta J$, since
$S^n=0$ for the $n \times n$ shift matrix \eqref{eq:S}.
\end{description}

\subsection{Model 2}
\label{sec:model2_numerical}

\begin{figure}[htbp]
\hskip -0.5cm
    \begin{tabular}{cc}
      \begin{minipage}[t]{0.47\hsize}
        \centering
        \includegraphics[keepaspectratio, scale=0.17]
        {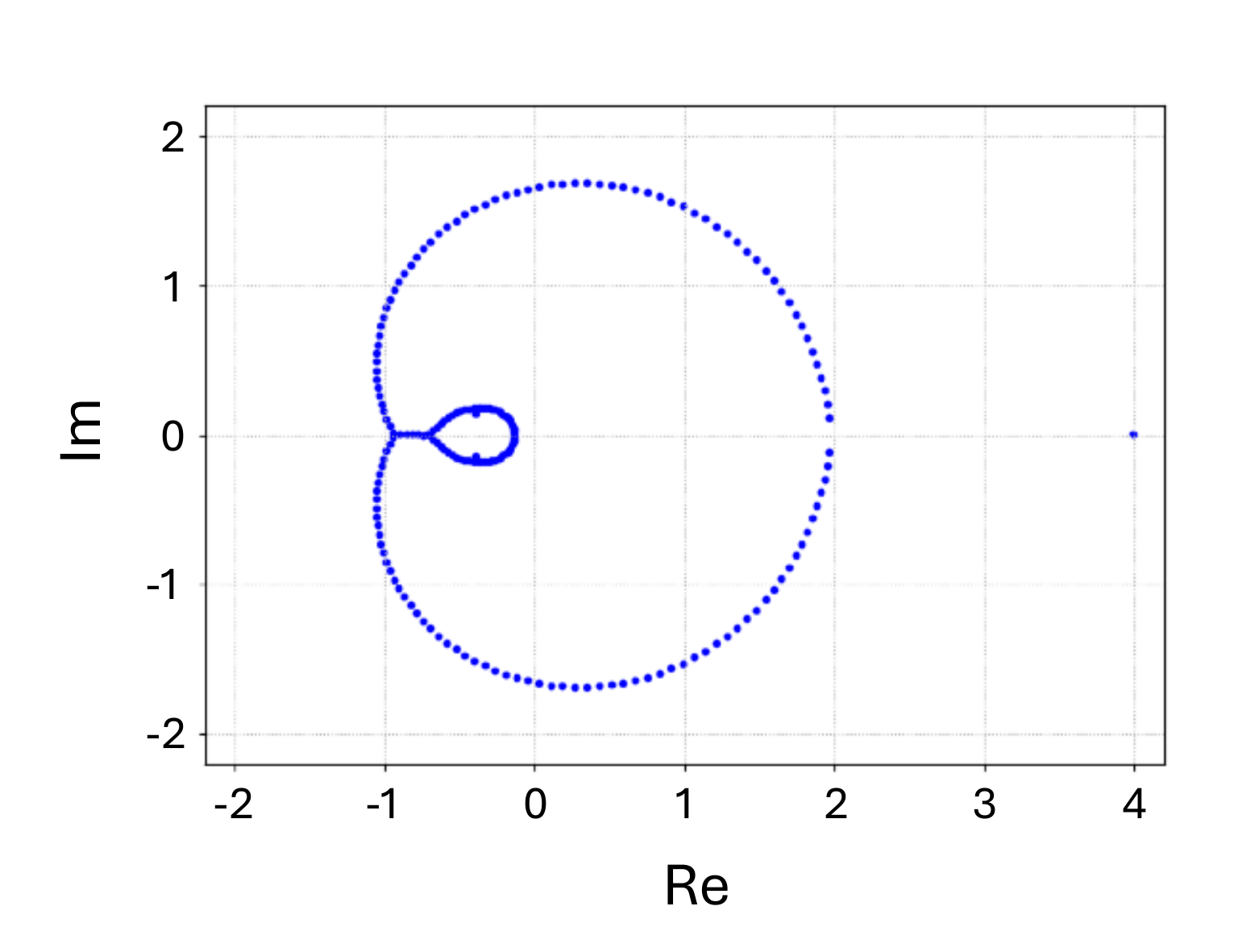}
        \subcaption{$m=1$}
        \label{fig:model2_numerical_m1}
      \end{minipage} &
      \begin{minipage}[t]{0.47\hsize}
        \centering
        \includegraphics[keepaspectratio, scale=0.17]
        {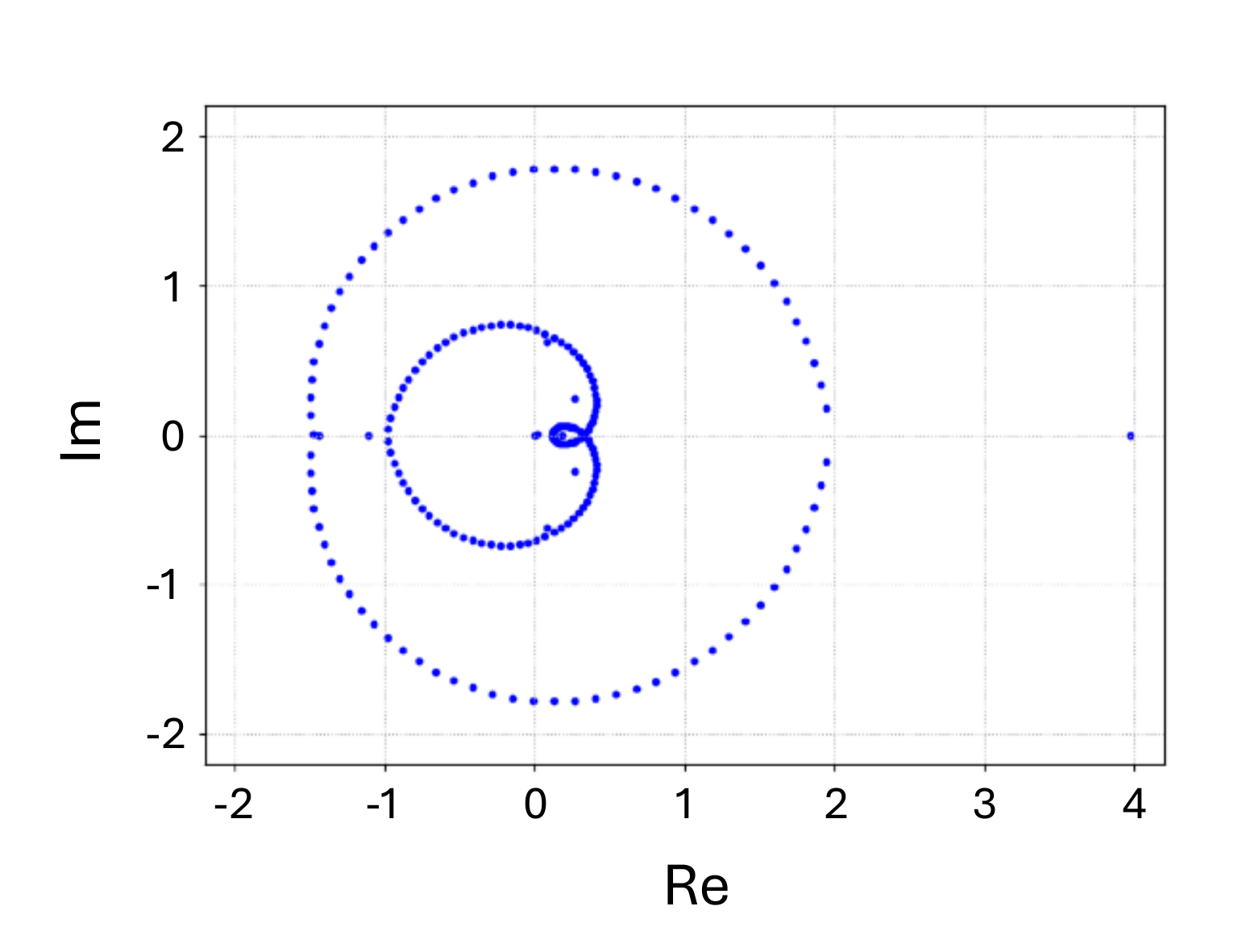}
        \subcaption{$m=2$}
        \label{fig:model2_numerical_m2}
      \end{minipage} \\
   
      \begin{minipage}[t]{0.47\hsize}
        \centering
        \includegraphics[keepaspectratio, scale=0.17]
        {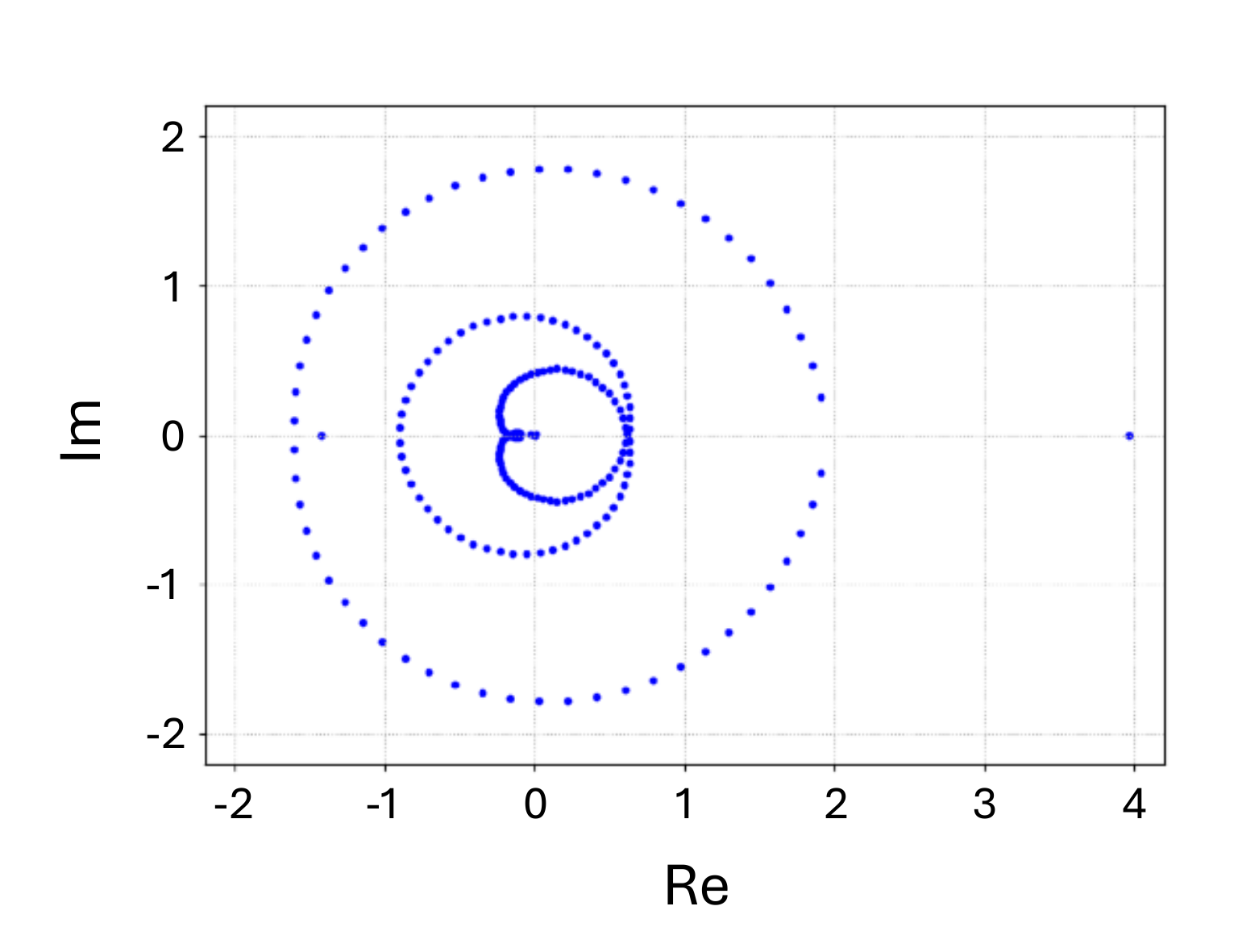}
        \subcaption{$m=3$}
        \label{fig:model2_numerical_m3}
      \end{minipage} &
      \begin{minipage}[t]{0.47\hsize}
        \centering
        \includegraphics[keepaspectratio, scale=0.17]
        {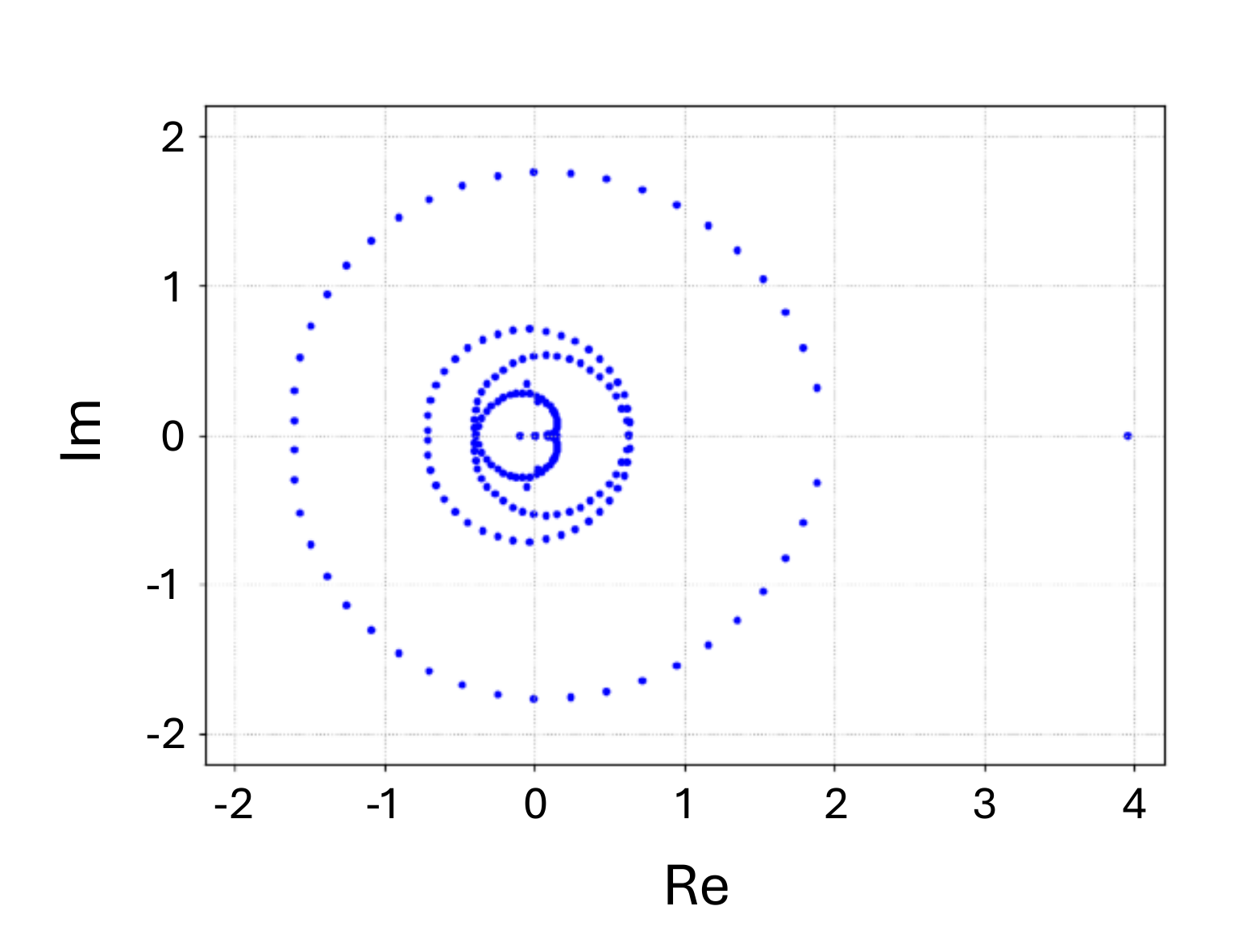}
        \subcaption{$m=4$}
        \label{fig:model2_numerical2_m4}
      \end{minipage} 
    \end{tabular}
     \caption{Numerically obtained eigenvalues are
     plotted for \textbf{model 2}, 
     $(S^{(2)}_{\delta J}(m))_{m=1}^n$, 
     with $n=200$, $\delta=0.01$, and $a=1$ 
     at time $m=1$, 2, 3, and 4, respectively.}
\label{fig:model2_numerical_ms}
\end{figure}

In Fig.~\ref{fig:model2_numerical_ms}, we dotted
the numerically obtained eigenvalues
of \textbf{model 2}
for $m=1,2,3$, and 4.
At $m=1$, a lima\c{c}on-like curve 
\cite{RT92,TE05} is observed.
At $m=2, 3$, and 4, 
a deformed circle whose radius is slightly less than 2
is formed by dots whose number decreases as
$m$ increases.
Notice that an outlier eigenvalue exists near $z=4$
in all figures.
The structure found in the vicinity of the origin
becomes more complicated as $m$ increases.
This inner structure shrinks rapidly to the origin 
when $m \geq 5$.
The motion of the outer dots is very
similar to that observed in \textbf{model 1}: 
They move along the upper- or lower-half deformed circles to $\R_-$, 
show repulsive motion on $\R_-$,
and then they are absorbed by the origin
one by one. 

\SSC{Exact Eigenvalue Processes}
\label{sec:ev}
\subsection{Model 1}
\label{sec:model1_ev}

For \textbf{model 1}, 
we consider the
following eigenvalue-eigenvector equations, 
\begin{equation}
S_{\delta J}^{(1)}(m) \v(m) = \lambda(m) \v(m),
\quad m=1,2,\dots, n.
\label{eq:ev1}
\end{equation}
Let $\1$ be the all-ones vector
and we introduce the Hermitian inner produce,
$\bra \u, \v \ket:= \sum_{j=1}^n u_j \overline{v_j}$,
$\u, \v \in \C^n$. 
Define
\[
\alpha(m) 
:= \bra \v(m), \1 \ket
=\sum_{j=1}^n v_j(m). 
\]
Let $1_{(\omega)}$ be the indicator function
of the condition $\omega$;
$1_{(\omega)}=1$ if $\omega$ is satisfied,
and $1_{(\omega)}=0$ otherwise.
The following fact will be used.
\begin{lem}
\label{thm:basic}
For $\ell \in \N$, 
\[
\bra S^{\ell} \1, \1 \ket
=(n-\ell) 1_{(1 \leq \ell \leq n-1)}.
\]
\end{lem}

For $m \in \{1, 2, \dots, n\}$, define
\begin{equation}
\widehat{p}(n, m) := \frac{n}{m}
\quad \mbox{and} \quad
p(n, m):=[ \widehat{p}(n, m) ],
\label{eq:p_hat}
\end{equation}
where $[x]$ denotes the greatest integer 
less than or equal to $x \in \R$
(the floor function of $x$).
Let
\begin{equation}
p_1 :=p(n-1, m).
\label{eq:p}
\end{equation}
The following is proved.

\begin{thm}
\label{thm:ev1}
For $m \in \{1, 2, \dots, n\}$, 
there are $p_1+1$ non-zero eigenvalues, which solve 
the following equation, 
\begin{align}
&\frac{1}{n \delta} z^{p_1+1}
- \frac{1-z^{p_1+1}}{1-z} 
\nonumber\\
& \quad 
+\frac{m}{n} \frac{1}{1-z}
\left\{ p_1+1 - \frac{1-z^{p_1+1}}{1-z} \right\}=0.
\label{eq:eigenvalues1}
\end{align}
This equation is also written as 
the polynomial equation, 
\begin{equation}
z^{p_1+1}
-n \delta \sum_{k=0}^{p_1} \left\{
1-(p_1-k) \frac{m}{n} \right\} z^k =0.
\label{eq:eigenvalues2}
\end{equation}
The corresponding eigenvectors satisfy $\alpha(m) \not=0$. 
All other $n-p_1-1$ eigenvalues degenerate at zero.
In this case, 
the corresponding eigenvectors satisfy
$\alpha(m)=0$, that is, they are orthogonal to $\1$.
\end{thm}
\vskip 0.3cm
\noindent{\it Proof} \,
Equation \eqref{eq:ev1} is written as
\begin{equation}
    (zI -S^m) \v(m) = \delta \alpha(m) \1, 
\label{eq:zI1}
\end{equation}
for $z=\lambda(m)$, 
where we noticed the equality, 
\begin{equation}
J \v(m) = \alpha(m) \1. 
\label{eq:Jv=a1}
\end{equation}
If we consider the zero-eigenvalue $z=\lambda_0=0$,
\eqref{eq:zI1} becomes $-S^m \v(m)=\delta \alpha(m) \1$.
Since $S^m$ shifts the elements of any vector upward by $m$
when $S^m$ is operated on the vector from the left,
the last $m$ elements of the vector $-S^m \v(m)$ 
are zero. Since $\1$ is the all-ones vector, 
$\alpha(m)$ should be 0.
For non-zero eigenvalues $z=\lambda(m) \not=0$ 
on the other hand, 
we can assume $\alpha(m) \not=0$. 
We solve this equation as follows,
\begin{align*}
\v(m) &= \delta \alpha(m) (z I -S^m)^{-1} \1
\nonumber\\
&= \delta \alpha(m) \sum_{k=0}^{\infty} z^{-(k+1)} S^{mk} \1,
\end{align*}
where we used the expansion formula of an inverse matrix.
By taking inner products with $\1$ on both sides, we have
\[
\bra \v(m), \1 \ket 
=\delta \alpha(m) \sum_{k=0}^{\infty} 
z^{-(k+1)} \bra S^{mk} \1, \1 \ket. 
\]
Since 
$mk \leq n-1, m,k \in \N 
\iff 
    k \leq p_1, 
    k \in \N, 
$
where $p_1$ is defined by \eqref{eq:p},
Lemma \ref{thm:basic} 
gives
\begin{align}
\alpha(m) =\delta \alpha(m) 
\sum_{k=0}^{p_1} z^{-(k+1)} (n-mk). 
\label{eq:alpha_eq1}
\end{align}
Here the fact $\bra \v(m), \1 \ket=\alpha(m)$ was used.
For non-zero eigenvalues, we can
assume $\alpha(m) \neq 0$. Hence we have
\begin{align}
    &1 =\delta \sum_{k=0}^{p_1} z^{-(k+1)} (n-mk)
    \nonumber\\
    \iff 
&\frac{1}{n \delta} 
    - z^{-1} \sum_{k=0}^{p_1} z^{-k} 
    + \frac{m}{n} z^{-1}  \sum_{k=0}^{p_1} k z^{-k} =0. 
\label{eq:eigenvaluesB1}
\end{align}
Then we use the summation formulas, 
\begin{align}
    z^{-1} \sum_{k=0}^{p_1} z^{-k} 
    &=z^{-(p_1+1)} \frac{1-z^{p_1+1}}{1-z}, 
    \nonumber\\
    z^{-1} \sum_{k=0}^{p_1} k z^{-k} 
    &=\frac{z^{-(p_1+1)}}{1-z}
    \left\{ p_1+1 - \frac{1-z^{p_1+1}}{1-z} \right\}.
\label{eq:expansion}
\end{align}
Thus \eqref{eq:eigenvaluesB1} is written as
\begin{align}
& \frac{1}{n \delta} 
-z^{-(p_1+1)} \frac{1-z^{p_1+1}}{1-z}
\nonumber\\
& \quad 
+ \frac{m}{n} \frac{z^{-(p_1+1)}}{1-z}
\left\{ p_1+1 - \frac{1-z^{p_1+1}}{1-z} \right\}
=0.
\end{align}
By multiplying $z^{p_1+1}$, 
\eqref{eq:eigenvalues1}
is obtained. 
Using the formulas \eqref{eq:expansion},
\eqref{eq:eigenvalues2} is readily 
derived from \eqref{eq:eigenvalues1}. 
\qed

\vskip 0.3cm
\noindent
\textbf{Remark 1.} \, 
Fix $m \in \{1,2, \dots, n\}$.
Suppose $\alpha(m)=0$, that is,
$\v(m)$ is orthogonal to $\1$.
By \eqref{eq:Jv=a1}, 
the eigenvalue-eigenvector equation
\eqref{eq:ev1} for $S_{\delta J}^{(1)}(m)$ is reduced to
that for $S^m$,
\[
S^m \v(m)=\lambda(m) \v(m).
\]
Consider the vectors in the form
\[
\v^0 =(v^0_1, v^0_2, \cdots, v^0_m, 0, \cdots, 0)^{\sf T}
\, \mbox{with} \, 
\sum_{j=1}^m v^0_j=0.
\]
Such vectors make $m-1$ dimensional space
of eigenvectors associated with the zero-eigenvalue 
$\lambda(m)= \lambda_0 =0$.
This means that the geometric multiplicity of $\lambda_0$
is $m-1$. Since Theorem \ref{thm:ev1} implies that
the algebraic multiplicity of $\lambda_0$ is given by
$n-p_1-1$ and 
$m-1 < n-p_1-1$ for $2 \leq m \leq n-2$,
$\lambda_0$ is defective and $S_{\delta J}^{(1)}(m)$
is nondiagonalizable for $2 \leq m \leq n-2$.
Now we take one of the eigenvectors,
$\v^0=\v^0_q$, $q=1,2,\dots, m-1$, and set
\[
\v_{q, \ell}=(S^{\sf T})^{m \ell} \v^0_q,
\]
where integers $\ell \in \{0, 1, \dots, p_1\}$ 
are chosen so that
$\bra \v_{q, \ell}, \1 \ket=0$. 
For example, when $n=6$ and $m=3$, we have
$m-1=2$ vectors
\[
\v^0_1=(1, -1, 0, 0, 0, 0)^{\sf T},
\quad
\v^0_2=(1, 0, -1, 0, 0, 0)^{\sf T}.
\]
Then, with $p_1=[(6-1)/3]=1$ and $\ell \in \{0, 1 \}$, 
we obtain the four linearly independent vectors,
\begin{align*}
& \v_{1,0} := \v^0_1,
\quad
\v_{1,1} := (S^{\sf T})^3 \v^0_1
=(0,0,0,1,-1,0)^{\sf T},
\nonumber\\
& \v_{2,0} := \v^0_2,
\quad
\v_{2,1} := (S^{\sf T})^3 \v^0_2
=(0,0,0,1,0,-1)^{\sf T},
\end{align*}
which satisfy the orthogonality to $\1$ 
and span the \textit{generalized eigenspace} for
the zero-eigenvalue $\lambda(3)=\lambda_0=0$
with dimensions
$n-p_1-1=6-1-1=4$. 
When $n=5$ and $m=3$, on the other hand, we have also 
$m-1=2$ vectors
\[
\v^0_1=(1, -1, 0, 0, 0)^{\sf T},
\quad
\v^0_2=(1, 0, -1, 0, 0)^{\sf T}.
\]
In this case, 
$\v_{2,1} :=(S^{\sf T})^3 \v^0_2
=(0,0,0,1,0)^{\sf T}$
does not satisfy the orthogonality condition; 
$\bra \v_{2,1}, \1 \ket \not=0$.
Then we have the
$n-p_1-1=5-[(5-1)/3]-1=5-1-1=3$
dimensional generalized eigenspace 
spanned by
$\{\v_{1,0}, \v_{1,1}, \v_{2,0}\}$
for $\lambda_0$.
For more systematic study of the generalized eigenspaces
of the present models,
see \cite{MKS24b}. 
\vskip 0.3cm

Let $\T_r$ and $\D_r$ be the circle (one-dimensional torus)
and the open disk
centered at the origin with radius $r>0$, respectively;
$\T_r:=\{z \in \C; |z| = r\}$ and
$\D_r:=\{z \in \C; |z| < r\}$. 
We can prove the following by Rouch\'e's theorem
(see, for instance, \cite[Section 4.4]{AF03}).
\begin{prop}
\label{thm:Rouche}
\begin{description}
\item{\textup{(i)}} \, 
All $p_1+1$ non-zero eigenvalues, which are given by
the solutions of \eqref{eq:eigenvalues2}, 
lie inside $\T_{n\delta +1}$.

\item{\textup{(ii)}} \, 
Assume that $n \delta > 3+ 2 \sqrt{2}=5.82 \cdots$.
Then the quadratic equation,
$r^2-(n \delta+1) r + 2 n \delta=0$, 
has two real solutions, 
\[
r_{\pm}:=\frac{n \delta}{2}+\frac{1}{2}
\pm \frac{n \delta}{2} \sqrt{1- \frac{6}{n \delta}
+\frac{1}{(n \delta)^2} },
\]
where 
$r_+ = n \delta-1-2/(n \delta)+ \rO((n \delta)^{-2})$
and 
$r_-=2+2/(n \delta)+\rO((n \delta)^{-2})$
as $n \delta \to \infty$.
The following holds.

\item{\textup{(iia)}} \,
Only one eigenvalue exists in 
$\D_{n \delta+1} \setminus \D_{r_+}
=\{z \in \C; r_+ \leq |z| < n \delta +1\}$.

\item{\textup{(iib)}} \,
There is no eigenvalue in 
$\D_{r_+} \setminus
(\D_{r_-} \cup \T_{r_-})
=\{z \in \C; r_- < |z| < r_+\}$.

\item{\textup{(iic)}} \,
Other $p_1$ non-zero eigenvalues lie in
$\D_{r_-} \cup \T_{r_-}
=\{z \in \C; |z| \leq r_- \}$.
\end{description}
\end{prop}
\vskip 0.3cm
\noindent{\it Proof} \,
\begin{description}
\item{(i)} \,
We set $f(z)=z^{p_1+1}$, 
\begin{align*}
g(z) &= - n \delta 
\sum_{k=0}^{p_1} \left\{1-(p_1-k)\frac{m}{n}\right\}
z^k. 
\end{align*}
For $k \in \{0,1,\dots, p_1\}$, we see that 
\[
\frac{1}{n} \le \left| 1-(p_1-k)\frac{m}{n} \right| \le 1.
\]
On $\T_{n\delta+1}$, $|f(z)| =
(n\delta+1)^{p_1+1}$ and 
\begin{align*}
 |g(z)| &\le n \delta \sum_{k=0}^{p_1} |z|^k =
n \delta \sum_{k=0}^{p_1} (n\delta+1)^k 
\nonumber\\
&=
(n\delta+1)^{p_1+1}-1
< |f(z)|. 
\end{align*}
By Rouch\'e's theorem, 
the numbers of zeros of $f(z)$ and $f(z) + g(z)$ inside
$\T_{n\delta+1}$ are the same. 
Therefore, the assertion is proved. 

\item{(ii)} \,
\noindent (ii)
Next we set $f(z) = - n \delta z^{p_1}$, 
\begin{align*}
g(z) &= z^{p_1+1} - n \delta \sum_{k=0}^{p_1-1}
\left\{1-(p_1-k)\frac{m}{n}\right\} z^k. 
\end{align*}
Assume $r > 1$. Then on $\T_r$ we see that
\begin{align*}
 |g(z)| &\le r^{p_1+1}+ n\delta \sum_{k=0}^{p_1-1}
  r^k 
  \nonumber\\
  &= n \delta \left(\frac{r}{n\delta} +
  \frac{1}{r-1}\right)r^{p_1} - \frac{n\delta}{r-1}. 
\end{align*}
If
\begin{equation}
\frac{r}{n \delta} + \frac{1}{r-1} \leq 1
\iff 
r^2-(n \delta +1) r + 2 n \delta \leq 0,
\label{eq:conditionA}
\end{equation}
then $|g(z)| < |f(z)|$, and thus 
the number of solution of \eqref{eq:eigenvalues2}
inside $\T_r$ is $p_1$. 
If $n \delta > 3 + 2 \sqrt{2}$, then
$\cD:=(n \delta+1)^2-8 n \delta > 0$
and $r_{\pm}=(n \delta+1 \pm \sqrt{\cD})/2 \in \R$
with $r_- < r_+$.
Hence, if $r_- < r < r_+$,
then \eqref{eq:conditionA} is satisfied.
It is easy to verify that $1 < r_- < r_+ < n \delta$.
Thus the assertions (iia)--(iic) are proved.
\qed
\end{description}
\vskip 0.3cm

We write the outlier eigenvalue specified by
Proposition \ref{thm:Rouche} (iia) 
as $\lambda_1(m)$.
Let $C_k, k \in \N_0$ be
the Catalan numbers \eqref{eq:Catalan} \cite{Rio68}. 

\begin{prop}
\label{thm:ev2}
\begin{description}
\item{\textup{(i)}} \,
If $p_1=0 \iff m=n$, then 
$\lambda_1(n)=n \delta$.

\item{\textup{(ii)}} \,
If $p_1 \geq 1 \iff m \leq n-1$, then, for $n \delta > 1$, 
we have the expression 
\eqref{eq:outlierA}. 
\end{description}
\end{prop}
\vskip 0.3cm
\noindent{\it Proof} \,
\begin{description}
\item{(i)} \,
\noindent (i) 
When $m=n$, that is $p_1=0$, 
\eqref{eq:eigenvalues2} is reduced
to $z-n \delta=0$. 

\item{(ii)} \,
\noindent (ii) 
When $p_1 =1$, \eqref{eq:eigenvalues2}
becomes the quadratic equation
\begin{equation}
z^2- n \delta z - n \delta \left( 1 - \frac{m}{n} \right)=0,
\label{eq:p1a}
\end{equation}
and we find
\begin{align*}
\lambda_1(m)&=\frac{n \delta}{2} \left[
1+\sqrt{1+\frac{4}{n \delta} \left(1-\frac{m}{n} \right) } \right]
\nonumber\\
&=n \delta+1 -\frac{m}{n} + 
\rO((n \delta)^{-1}).
\end{align*}
When $p_1 \geq 2$, the assertion is proved by
induction: 
For $q=1, 2, \dots, p_1-1$, we assume
\[
z=n\delta+1
- \sum_{k=0}^{q-1} C_k 
\frac{(m/n)^{k+1}}{(n \delta)^k}
- c \left(\frac{m}{n} \right)^{q+1}
\frac{1}{(n \delta)^q}
\]
with unknown coefficient $c$.
Insert this into \eqref{eq:eigenvalues2}.
Then $c$ is determined to be the $q$-th Catalan number $C_q$.
\qed
\end{description}

\vskip 0.3cm
\noindent
\textbf{Remark 2.} \, 
\begin{description}
\item{(i)} \, 
\noindent (i) 
If we set $n=200$, $\delta=0.01$, and $m=1$,
\eqref{eq:outlierA} gives
\begin{align*}
\lambda_1(1) &= 2+1-\frac{1}{200}
-\frac{1}{200^2 \times 2} 
-\frac{2}{200^3 \times 2^2} -\cdots
\nonumber\\
&=2.994 \cdots.
\end{align*}
This implies the fact that a dot near $z=3$ 
in Fig.~\ref{fig:model1_numerical_m1}
shows an exact eigenvalue.
A unit circle with a gap at $z=1$
in this figure shall consist of
$p_1=[(200-1)/1]=199$ exact eigenvalues
as asserted by Theorem \ref{thm:ev1}. 

\item{(ii)} \, 
\noindent (ii) 
When $n=200$, we have
$p_1=[(n-1)/m]=99, 24, 13$, and 2 for
$m=2, 8, 15$, and 80, respectively.
Hence the dots on the outer circle
in Fig.~\ref{fig:model1_numerical_m2}
and the dots on the outer circle 
in Fig.~\ref{fig:model1_numerical_m8}
are exact eigenvalues.
The 13 and the 2 dots, 
which are not equal to zero in
Fig.~\ref{fig:model1_numerical_m15}
and Fig.~\ref{fig:model1_numerical_m80}, respectively, 
represent exact eigenvalues.
Notice that in these figures the outlier eigenvalue 
$\lambda_1(m) \simeq 3$
is out of the frames.
All other $n-p_1-1$ eigenvalues are degenerated
at the origin. 

\item{(iii)} \, 
\noindent (iii) 
The dots forming the inner circle
with radius $\simeq 0.7$ 
and the two non-zero dots near the origin
in Fig.~\ref{fig:model1_numerical_m2},
the annulus with wavy boundaries 
and the two non-zero dots near the origin
in Fig.~\ref{fig:model1_numerical_m8}, 
and the small annulus surrounding the origin
in Fig.~\ref{fig:model1_numerical_m15}
are all not exact eigenvalues of \textbf{model 1},
but shall be eigenvalues of the system
perturbed by rounding errors of computer.
They represent structures of pseudospectrum
including $\lambda_0$ of \textbf{model 1}.
\end{description}

\vskip 0.3cm
\noindent
\textbf{Remark 3.} \, 
If 
$1 \leq n/m < 2 \iff n/2 < m \leq n$,
then $p_1=1$. 
Assume that $\delta \in \R$.
In this case, \eqref{eq:eigenvalues2} 
becomes the quadratic equation
\eqref{eq:p1a}.
Put $z=x+iy$, $x, y \in \R$. Then we obtain the
following equations from \eqref{eq:p1a},
\begin{align}
& x^2-y^2 - 
n \delta \left\{ x +\left(1- \frac{m}{n} \right) \right\}=0,
\nonumber\\
& y ( 2x- n \delta )=0.
\label{eq:p1b}
\end{align}
The second equation 
in \eqref{eq:p1b} gives
$y=0$ or $x=n\delta/2$.
If we assume $x=n \delta/2$, 
then the first equation in
\eqref{eq:p1b} gives
$y^2=- (n\delta)^2/4
-n \delta (1-m/n)$. 
Since $1-m/n \geq 0$, the RHS 
is negative, and thus 
this contradicts $y \in \R$.
Hence $y =0$. 
Then the first equation in
\eqref{eq:p1b} becomes
$x^2 - n \delta x - n \delta (1-m/n)=0$, 
which is solved by
$x_{\pm}:= (n \delta/2) \{
1 \pm \sqrt{1 + (4/(n \delta)) (1-n/m)} \}$.
When $n \delta \gg 1$, 
$x_{+} \simeq n \delta + 1 - m/n$ and
$x_{-} \simeq -1 +m/n$. 
Hence $x_+$ should be identified with $\lambda_1(m)$.
We see that
\begin{align*}
\lim_{m \searrow n/2} x_- &=
-(n \delta/2) \{ 
\sqrt{1+2/(n \delta)} -1 \}
=: x_-^0 <0,
\nonumber\\
\lim_{m \nearrow n} x_- &=0.
\end{align*}
Therefore, the non-zero eigenvalue 
$x_-=x_-(m)$,
which is not the outlier $\lambda_1(m)$, 
moves from $x_-^0$ to 0 on $\R_-$ as $m$ increases
from $n/2$ to $n$.

\subsection{Model 2}
\label{sec:model2_ev}

\begin{figure}[htbp]
    \begin{tabular}{cc}
      \begin{minipage}[t]{0.47\hsize}
        \centering
        \includegraphics[keepaspectratio, scale=0.27]
        {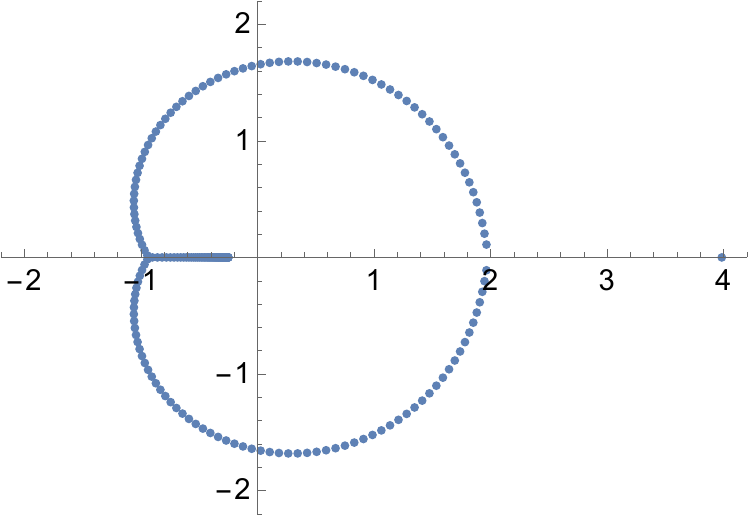}
        \subcaption{$m=1$}
        \label{fig:model2_exact_m1}
      \end{minipage} &
      \begin{minipage}[t]{0.47\hsize}
        \centering
        \includegraphics[keepaspectratio, scale=0.27]
        {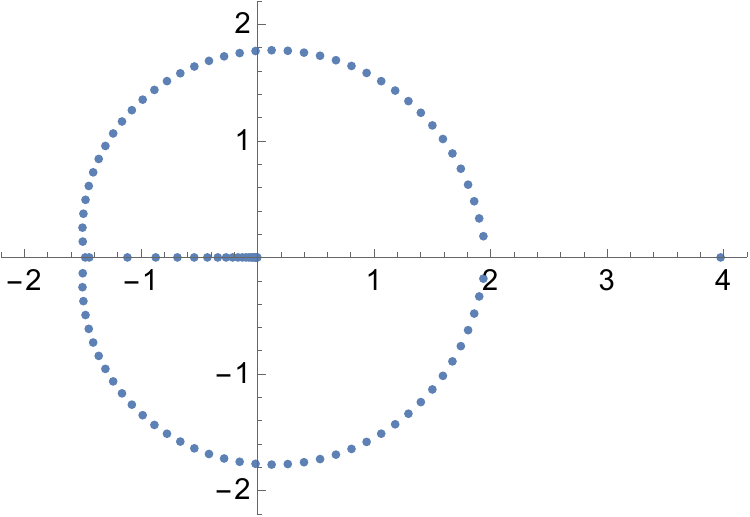}
        \subcaption{$m=2$}
        \label{fig:model2_exact_m2}
      \end{minipage} \\
   
      \begin{minipage}[t]{0.47\hsize}
        \centering
        \includegraphics[keepaspectratio, scale=0.27]
        {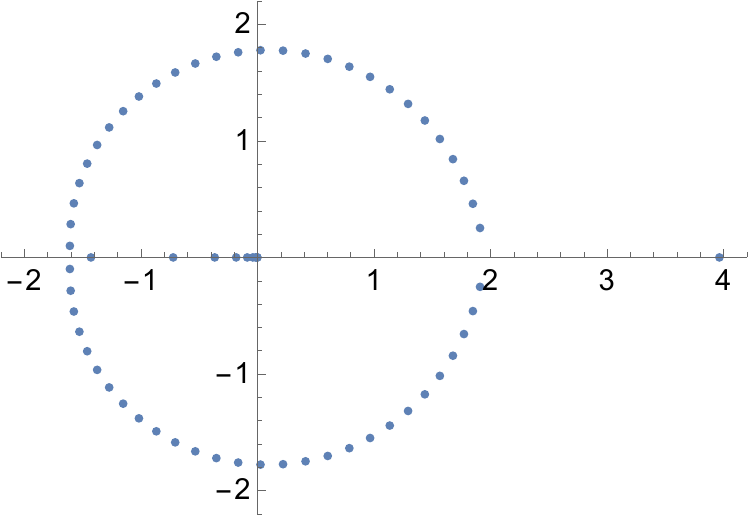}
        \subcaption{$m=3$}
        \label{fig:model2_exact_m3}
      \end{minipage} &
      \begin{minipage}[t]{0.47\hsize}
        \centering
        \includegraphics[keepaspectratio, scale=0.27]
        {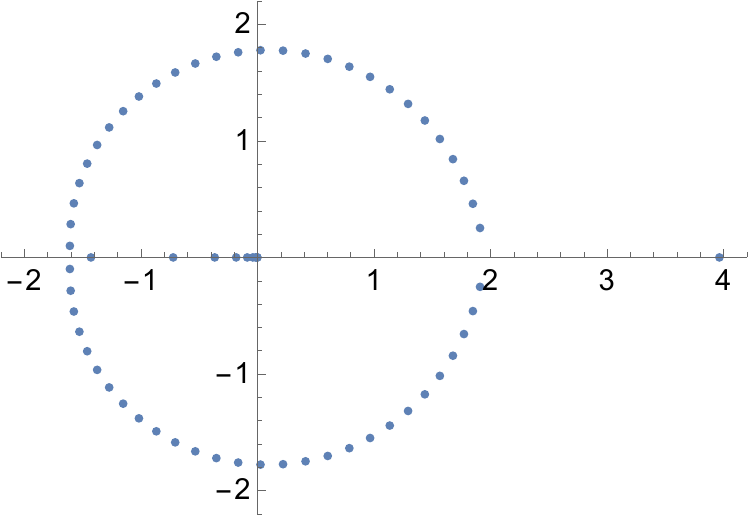}
        \subcaption{$m=4$}
        \label{fig:model2_exact_m4}
      \end{minipage} 
    \end{tabular}
     \caption{Exact eigenvalues are
     plotted for \textbf{model 2}, 
     $(S^{(2)}_{\delta J}(m))_{m=1}^n$, 
     with $n=200$, $\delta=0.01$, and $a=1$
     at $m=1, 2, 3$, and 4, respectively.}
\label{fig:model2_exact_ms}
  \end{figure}

For \textbf{model 2}, 
we consider the eigenvalue problem, 
\[
S^{(2)}_{\delta J}(m) \v(m) = \lambda(m) \v(m). 
\]
Let
\begin{equation}
p_2=p(n-1, m+1),
\label{eq:p2}
\end{equation}
where $p(n, m)$ was defined by \eqref{eq:p_hat}.
Then Theorem \ref{thm:ev1} is
generalized as follows \cite{MKS24b}. 
\begin{thm}
\label{thm:evG1}
For $m \in \{1, 2, \dots, n\}$, 
there are $p_1+1$ non-zero eigenvalues, which solve 
the following equation, 
\begin{align}
&\frac{1+a}{n \delta} 
\left(\frac{z}{1+a} \right)^{p_1+1}
- \frac{1-\{z/(1+a)\}^{p_1+1} }
{1-z/(1+a)}
\nonumber\\
& \quad
+ [
m/n + a/\{(1+a)n\} ]
\frac{1}{1-z/(1+a)}
\nonumber\\
& \qquad \times
\left[
p_1+1
- 
\frac{1-\{z/(1+a)\}^{p_1+1} }
{1-z/(1+a) }
\right]
\nonumber\\
& \quad
- 1_{(p_1 \geq p_2+1, \, p_1 \geq (n+1)/(m+1))} \frac{1}{(1+a)^{p_1}}
\nonumber\\
& \qquad \times
\sum_{k=0}^{p_1-p_2-1} z^k
\sum_{q=n-m(p_1-k)+1}^{p_1-k} a^q
\binom{p_1-k}{q}
\nonumber\\
& \qquad \times [ q-\{n-m(p_1-k)\}]/n
=0.
\label{eq:eigenvaluesG1}
\end{align}
This equation is also written as the polynomial equation
\begin{align}
&
\left( \frac{z}{1+a} \right)^{p_1+1}
- \frac{n \delta}{1+a}
\nonumber\\
& \, \,
\times \sum_{k=0}^{p_1} \left[
1-  (p_1-k) 
\left\{
\frac{m}{n} + \frac{a}{(1+a) n}
\right\}
\right]
\left(\frac{z}{1+a} \right)^k
\nonumber\\
& \,
- 1_{(p_1 \geq p_2+1, \, p_1 \geq (n+1)/(m+1))} \frac{n \delta}{(1+a)^{p_1+1}}
\nonumber\\
& \quad
\times
\sum_{k=0}^{p_1-p_2-1} z^k
\sum_{q=n-m(p_1-k)+1}^{p_1-k} a^q
\binom{p_1-k}{q}
\nonumber\\
& \quad
\times \frac{1}{n} [ q-\{n-m(p_1-k)\}]
=0.
\label{eq:eigenvaluesG2}
\end{align}
The corresponding eigenvectors satisfy $\alpha(m) \not=0$. 
All other $n-p_1-1$ eigenvalues degenerate at zero.
In this case, 
the corresponding eigenvectors satisfy
$\alpha(m)=0$, that is, they are orthogonal to $\1$.
\end{thm}
As a matter of course, if we put $a=0$,
\eqref{eq:eigenvaluesG1} and 
\eqref{eq:eigenvaluesG2} are reduced to
\eqref{eq:eigenvalues1} and \eqref{eq:eigenvalues2},
respectively. 

\vskip 0.3cm
\noindent
\textbf{Remark 4.} \,
Consider the difference in $m$ of $\widehat{p}$, which is
given by \eqref{eq:p_hat} as
\[
\Delta \widehat{p} := 
\widehat{p}(n-1, m) - \widehat{p}(n-1, m+1)
=\frac{n-1}{m(m+1)}.
\]
Here we regard $\Delta \widehat{p}$ as a function
of the real variable $m \geq 1$, and for $s \not= 0$
we solve the equation
\begin{align*}
\Delta \widehat{p} = s
\quad &\iff \quad
m^2+m-\frac{n-1}{s}=0.
\end{align*}
Let $m(n, s)$ be the positive solution,
\begin{equation}
m(n, s) := \sqrt{ \frac{n-1}{s}+\frac{1}{4} } - \frac{1}{2}. 
\label{eq:m_n_s}
\end{equation}
By definition, if $1 \leq m < m(n, s)$, then
$\Delta \widehat{p} > s$
$\Longrightarrow p_1-p_2 \gtrsim s$,
and if $m(n, 1/s) < m$,  then
$\Delta \widehat{p} < 1/s$
$\Longrightarrow p_1-p_2 \lesssim 1/s$.
Suppose that $s \in \N$.
Then the above calculation will be interpreted as follows:
When $m \simeq m(n, s)$, 
the number of terms in the last part of the 
left-hand-side of \eqref{eq:eigenvaluesG1} 
is about $s$. And
at about $s$ successive values of $m$ 
around $m(n, 1/s)$, $p_1=p_2$, 
and hence the last part of the left-hand-side
of \eqref{eq:eigenvaluesG1} vanishes.
For example, when $n=10^5$, 
$m(10^5, 10) = 99.50 \cdots$,
$m(10^5,1/10) = 999.4 \cdots$, and we can see that
$p_1-p_2=999-990=9$ at $m=100$,
and that
$p_1=p_2=99$ for 
the ten values of $m$; 
$m=1000, 1001, \cdots, 1009$.
\vskip 0.3cm
The dependence of $m(n, s)$ on $n$ expressed by
$\sqrt{n-1}$ in \eqref{eq:m_n_s} is essential,
and the following lemma is valid.
We write the smallest integer greater than or equal to
$x \in \R$ as $\lceil x \rceil$ (the ceiling function of $x$). 
Remark that the floor function of $x$ is
denoted by $[x]$ in this paper.
\begin{lem}
\label{thm:p_0_1}
Let
$I_{n-1}:=[ \lceil \sqrt{n-1} \, \rceil, n-1 ] \cap \N$ 
and
$T_{n-1}:=\{ [(n-1)/k] ; k=1,2, \dots, n-1\}$.
Then
\[
p_1-p_2=\begin{cases}
1, \quad \mbox{if $m \in I_{n-1} \cap T_{n-1}$},
\cr
0, \quad \mbox{if $m \in I_{n-1} \setminus T_{n-1}$}.
\end{cases}
\]
\end{lem}
Proposition \ref{thm:ev2} for \textbf{model 1} is 
generalized for \textbf{model 2} as follows.
\begin{prop}
\label{thm:evG2}
We have an outlier eigenvalue 
$\lambda_1(m)$, whose modulus 
goes to $\infty$ as $n \delta \to \infty$.
The following holds for $m \in \{1, 2, \dots, n\}$.
\begin{description}
\item{\textup{(i)}} \,
If $p_1=0 \iff m=n$, then 
$\lambda_1(n)=n \delta$.

\item{\textup{(ii)}} \,
If $p_1 \geq 1 \iff m \leq n-1$, then, for $n \delta >1$,
we have the expression 
\begin{align}
& \lambda_1(m) =n \delta+ 1 +a
\nonumber\\
& \, -
(1+a) \sum_{k=0}^{p_1-1} C_k 
\left( \frac{m}{n}+\frac{a}{(1+a) n}
\right)^{k+1} 
\nonumber\\
& \quad \times
\left( \frac{1+a}{n \delta} \right)^k
+\rO((n \delta)^{-p_1})
\nonumber\\
& \, 
+1_{(p_1 \geq p_2+1)} \rO((n \delta)^{-p_2}).
\label{eq:large_evG}
\end{align}
\end{description}
\end{prop}
\vskip 0.3cm
\noindent{\it Proof} \,
Comparing the left-hand-side of \eqref{eq:eigenvalues1}
for \textbf{model 1}
and the first three terms in the left-hand-side of
\eqref{eq:eigenvaluesG1} for \textbf{model 2}, 
we find that the latter is obtained from the former
by the following replacement,
\begin{equation}
z \to \frac{z}{1+a}, 
\, 
n \delta \to \frac{n \delta}{1+a},
\, 
\frac{m}{n} \to \frac{m}{n}+\frac{a}{(1+a)n},
\label{eq:replace1}
\end{equation}
associated with the introduction of the parameter $a$
in \textbf{model 2}.
The first three lines of \eqref{eq:large_evG} 
are obtained from
\eqref{eq:outlierA} by this replacement \eqref{eq:replace1}.
The correction $\rO((n \delta)^{-p_2})$ should be added
due to the last part in the left-hand-side of 
\eqref{eq:eigenvaluesG1}. \qed

\vskip 0.3cm
\noindent
\textbf{Remark 5.} \,
\begin{description}
\item{(i)} \,
\noindent (i) 
If we set $n=200$, $\delta=0.01$, and $a=1$,
\eqref{eq:large_evG} gives
\begin{align*}
\lambda_1(m) &\simeq 2+1+1
-2 \left( 
\frac{m}{200}+\frac{1}{2 \times 200} 
\right)
\nonumber\\
&=3.995 - \frac{m}{100}.
\end{align*}
This implies the fact that a dot near $z=4$ 
in each figure of Fig.~\ref{fig:model2_numerical_ms}
shows an exact eigenvalue. 

\item{(ii)} \, 
\noindent (ii) 
The exact eigenvalues given by
the solutions of \eqref{eq:eigenvaluesG1} 
of Theorem \ref{thm:evG1}
are plotted in Fig.~\ref{fig:model2_exact_ms}
for $m=1,2,3$, and 4.
Comparing Fig.~\ref{fig:model2_numerical_ms}
and Fig.~\ref{fig:model2_exact_ms}, 
the dots located in the outermost regions
are exact eigenvalues.
The exact eigenvalues located in the inner regions,
especially most of the exact eigenvalues 
on $\R_-$, are missing
in the numerical results. 
The patterns observed in the vicinity 
of the origin in the numerical results shown by
Fig.~\ref{fig:model2_numerical_ms} 
consist of eigenvalues of systems
perturbed by uncontrolled rounding errors of computer,
which visualize structures of pseudospectra
including $\lambda_0$ of $S_{\delta J}^{(2)}(m)$.
\end{description}
\vskip 0.3cm

\SSC{Pseudospectrum Processes}
\label{sec:pseudo}

Consider the \textit{banded Toeplitz matrices} such that
the number of diagonal lines in which
the elements are non-zero is finite
and given by $2w +1$, $w \in \N_0$.
Let $\{A_n \}$ be a family of 
such Toeplitz matrices with sizes 
$n \in \N$;
\[
A_n=( (A_n)_{jk} )_{1 \leq j, k \leq n}
=( a_{j-k} 1_{(|j-k| \leq w)} )_{1 \leq j, k \leq n}.
\]
We write the matrix representation of
the corresponding \textit{banded Toeplitz operator}
as
$\widehat{A}
=( a_{j-k} 1_{(|j-k| \leq w)} )_{j, k \in \N}$.
The \textit{symbol} of $\widehat{A}$ 
is defined as \cite{BS99,TE05} 
$f_{\widehat{A}}(z) := \sum_{\ell; |\ell| \leq w}
a_{\ell} z^{\ell}$.
Let $\T:=\T_1=\{e^{i \theta} ; \theta \in [0, 2 \pi)\}$, 
i.e., the unit circle.
Then the \textit{symbol curve} 
is defined by \cite{BS99,TE05}
\[
f_{\widehat{A}}(\T)
=\{f_{\widehat{A}}(z) ; z \in \T \}.
\]
Given a point $z \in \C \setminus f_{\widehat{A}}(\T)$,
$I(f_{\widehat{A}}, z)$ is defined 
as the \textit{winding number} of $f_{\widehat{A}}(\T)$ 
about $z$ in the usual positive (counterclockwise) sense.
The following theorem is well-known \cite{BS99,TE05}.
\begin{thm}
\label{thm:Toeplitz}
Let $\sigma(\widehat{A})$ be
the spectra of $\widehat{A}$.
Then $\sigma(\widehat{A})$ is equal to
$f_{\widehat{A}}(\T)$ together with all the points
enclosed by this curve with 
$I(f_{\widehat{A}},z) \not=0$.
\end{thm}
The following fact was proved \cite{RT92,TE05}.
\begin{prop}
\label{thm:pseudoS}
For some $M > 1$ and all sufficiently large $n$, 
\[
\| (zI-A_n)^{-1} \| \geq M^n
\quad 
\mbox{for any $z \in \sigma(\widehat{A})$}.
\]
\end{prop}
This implies that the pseudospectra
of $A_n$ will reflect 
the exact spectra of the corresponding
Toeplitz operator $\widehat{A}$.

\subsection{Model 1}
\label{sec:model1_pseudo}

\begin{figure}[htbp]
    \begin{tabular}{ccc}
      \begin{minipage}[t]{0.34\hsize}
        \centering
        \includegraphics[keepaspectratio, scale=0.17]
        {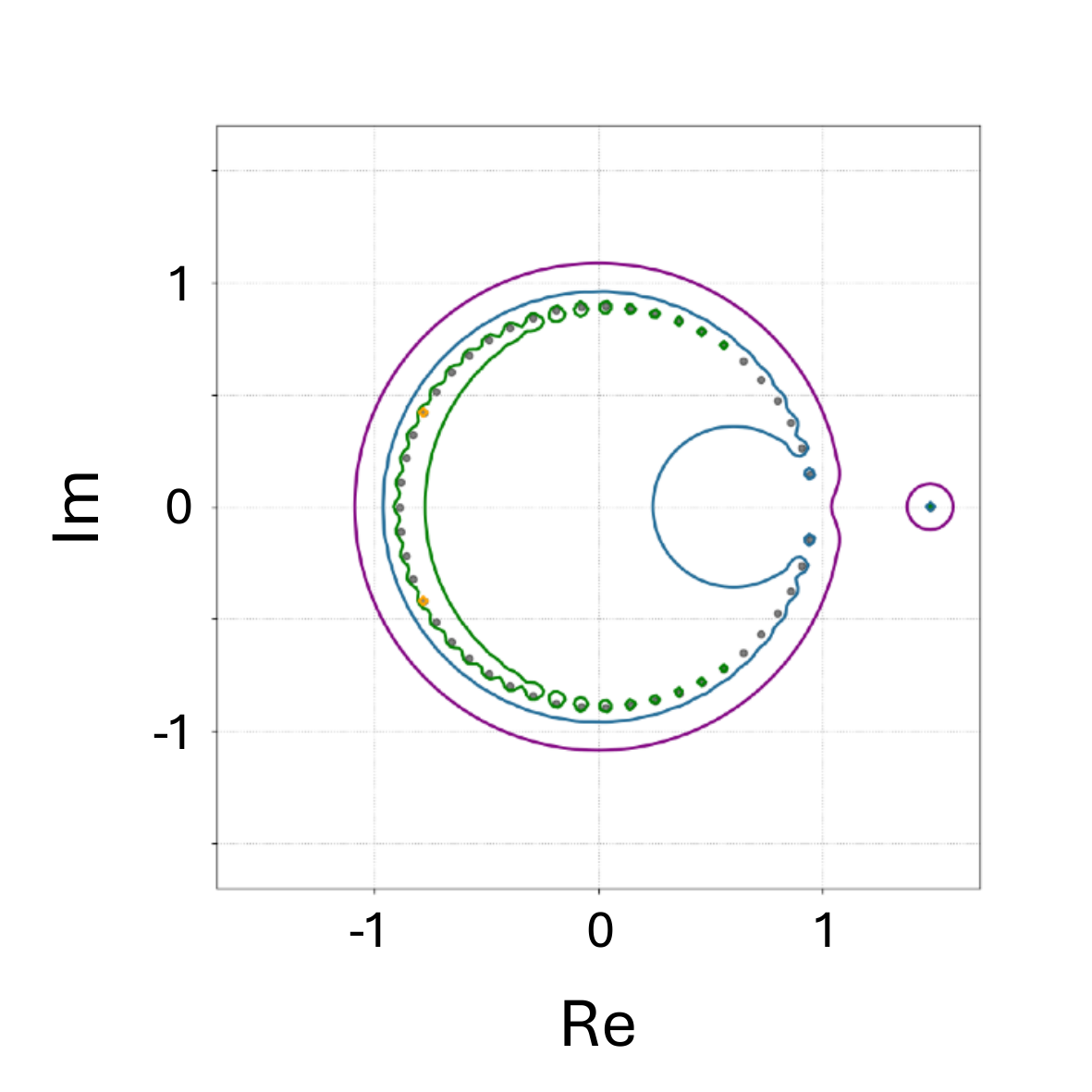}
        \subcaption{$m=1$}
        \label{fig:ps_m1}
      \end{minipage} &
      \begin{minipage}[t]{0.34\hsize}
        \centering
        \includegraphics[keepaspectratio, scale=0.17]
        {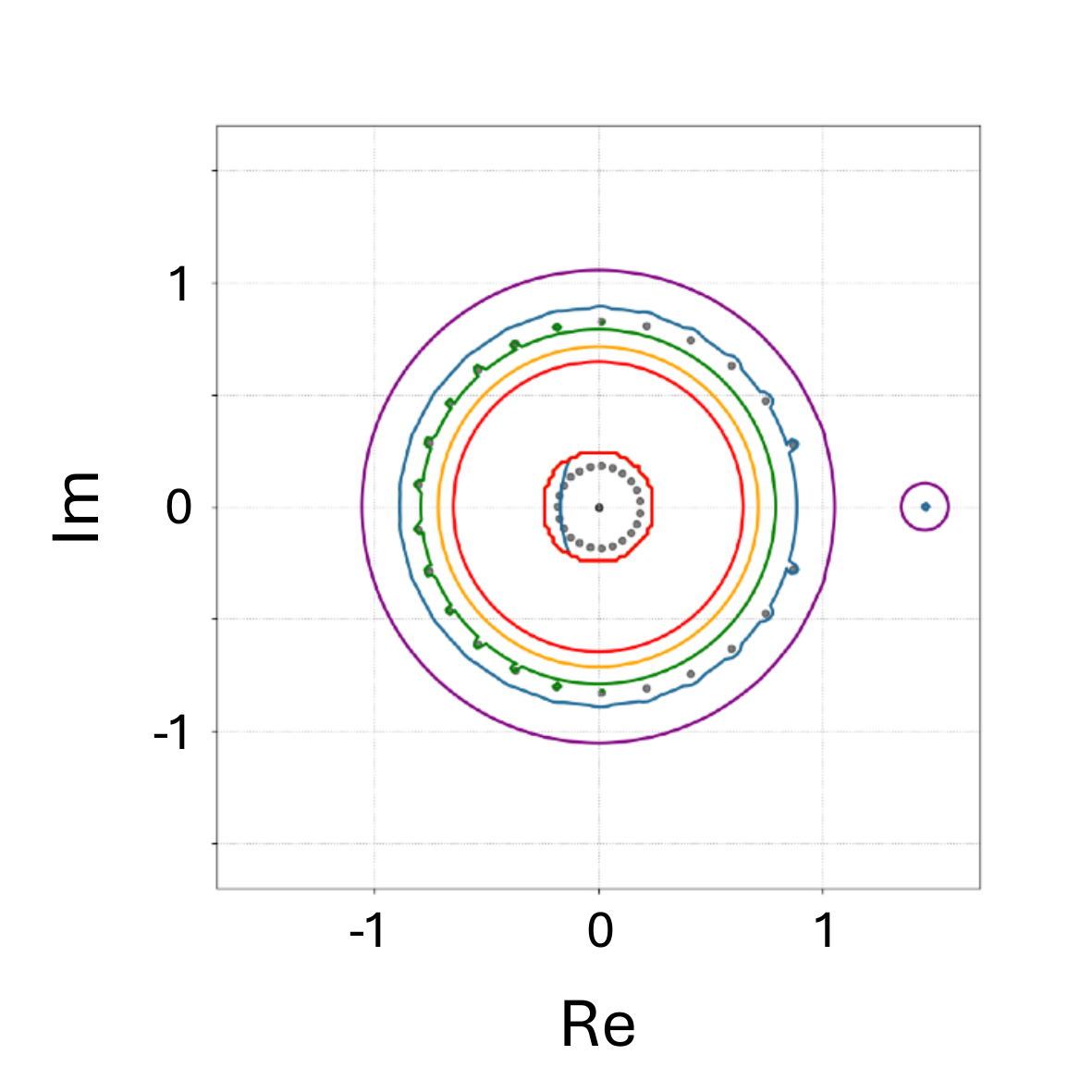}
        \subcaption{$m=2$}
        \label{fig:ps_m2}
      \end{minipage} &
      \\
      \begin{minipage}[t]{0.34\hsize}
        \centering
        \includegraphics[keepaspectratio, scale=0.17]
        {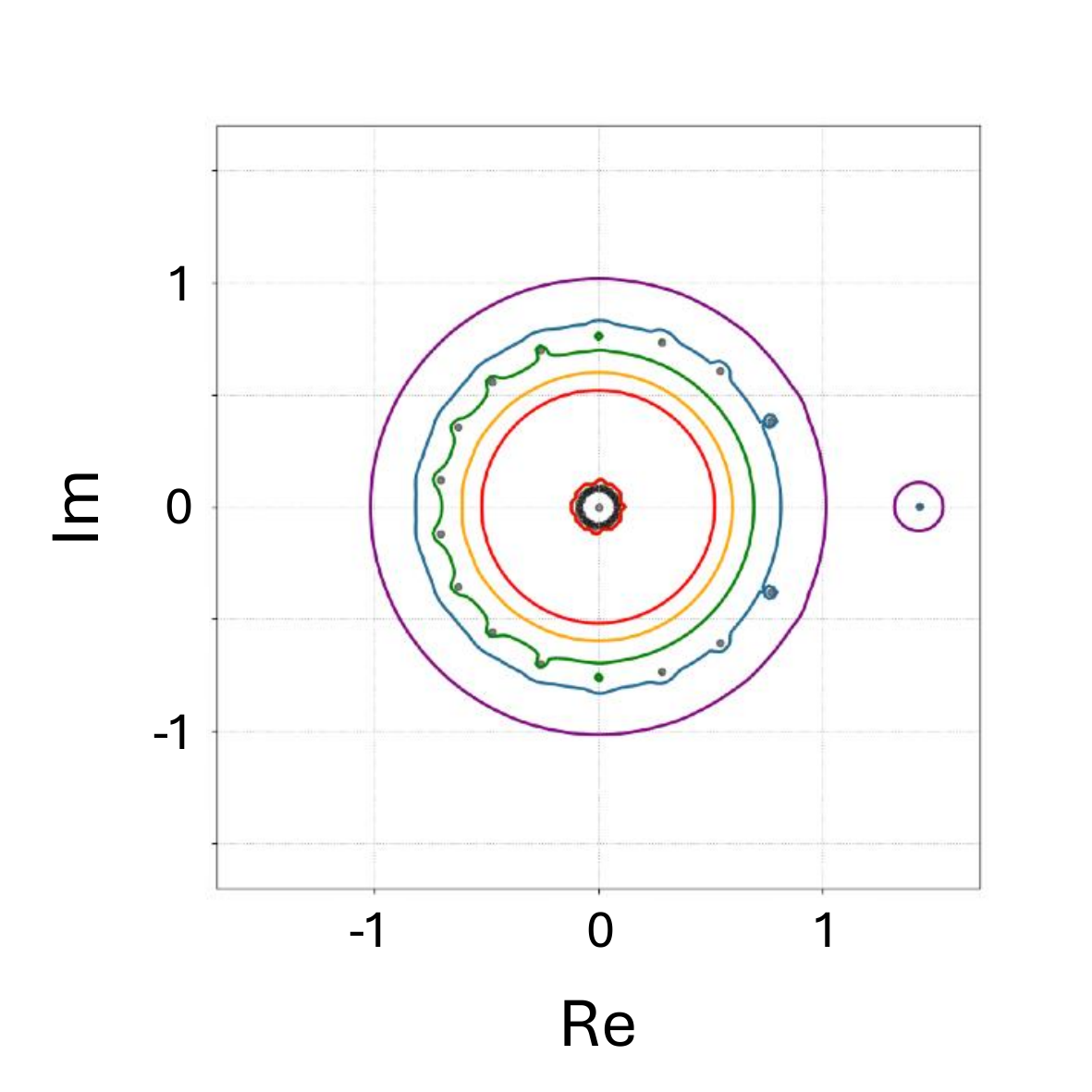}
        \subcaption{$m=3$}
        \label{fig:ps_m3}
      \end{minipage} &
      \begin{minipage}[t]{0.34\hsize}
        \centering
        \includegraphics[keepaspectratio, scale=0.17]
        {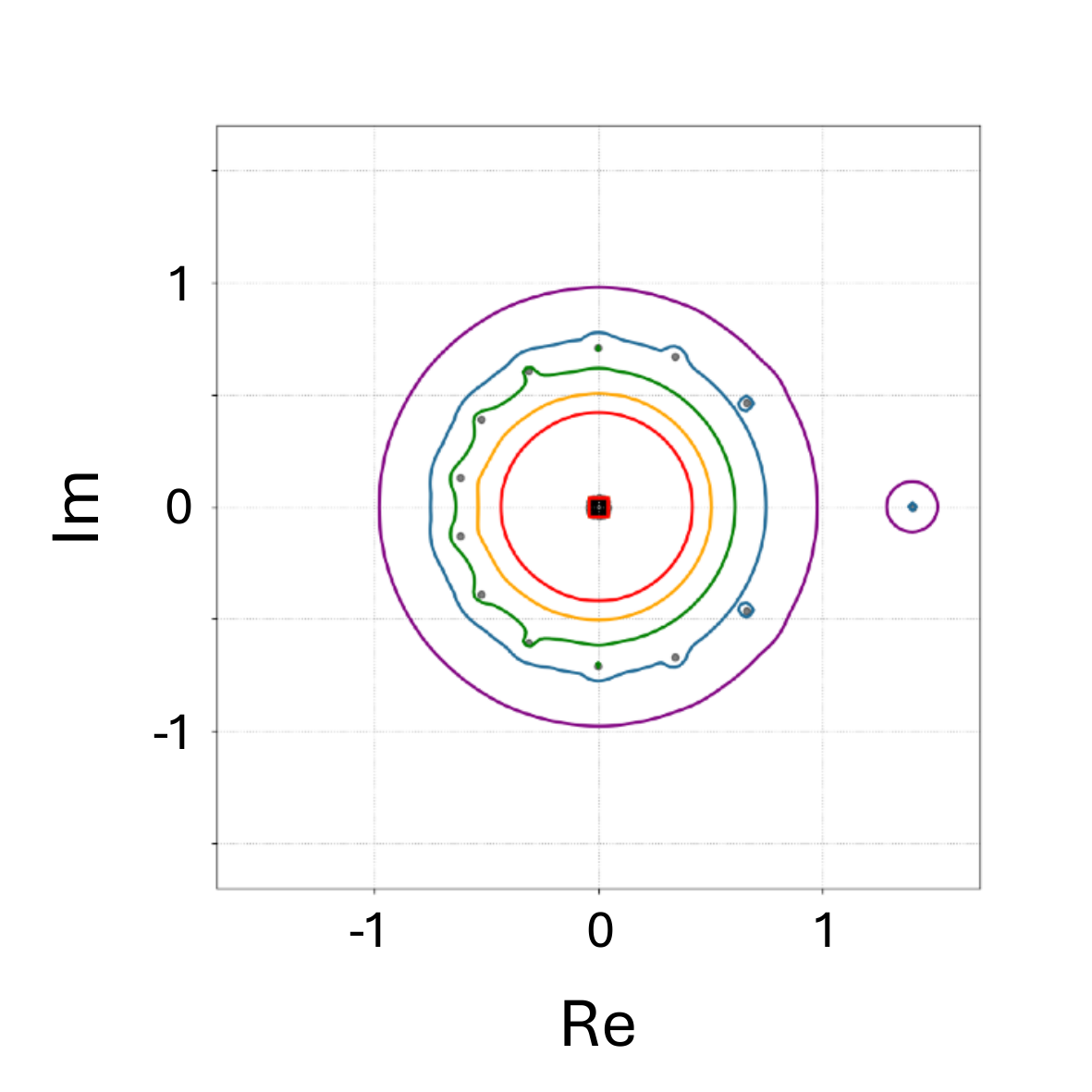}
        \subcaption{$m=4$}
        \label{fig:ps_m4}
      \end{minipage} &
      \begin{minipage}[t]{0.20\hsize}
        \centering
        \includegraphics[keepaspectratio, scale=0.12]
        {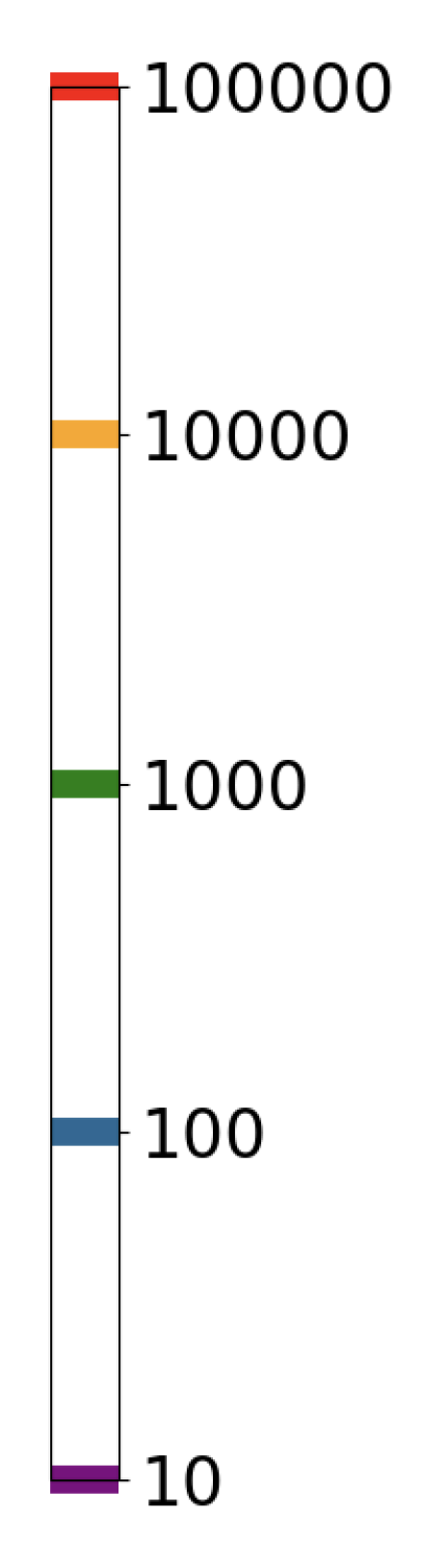}
        \subcaption{Color scale}
        \label{fig:scale}
      \end{minipage} 
    \end{tabular}
     \caption{Contour plots of
     $\|(zI-S^{(1)}_{\delta J}(m))^{-1}\|$
     for \textbf{model 1} with
     $n=50$ and $\delta=0.01$ 
     at $m=1,2,3$, and 4, respectively.}
\label{fig:ps}
  \end{figure}
  
The Toeplitz operators corresponding to
$S^m$, $m \in \{1,2, \dots, n \}$, are given by
\[
\widehat{S}^m=
\Big( (\widehat{S}^m)_{jk} \Big)_{j, k \in \N}
=\Big( \delta_{j \, k-m} \Big)_{j, k \in \N}, 
\quad m \in \N.
\]
The symbols of $\widehat{S}^m$,
$m \in \N$ are given by
$f_{\widehat{S}^m}(z) =z^m$.
Let $\D$ be the open unit disk,
$\D:=\D_1=\{z \in \C; |z| < 1\}$, 
and we write 
$\overline{\D}:=\{z \in \C; |z| \leq 1\}
=\D \cup \T$.
We see that
$f_{\widehat{S}^m}(\T)=\T$
and all points enclosed by $\T$ have
winding number $m \in \N \not=0$.
Hence by Theorem \ref{thm:Toeplitz}, 
$\sigma(\widehat{S}^m)=\overline{\D}$, $m \in \N$.

The present \textbf{model 1}
can be regarded as the system such that
the rank 1 perturbation $\delta J$ is added to
$S^m$, $m=1,2, \dots, n$.
It has been reported in many examples
(see, for instance, Fig.~7.4 and explanations in
Section 7 of \cite{TE05}),
caused by dense random perturbations 
the eigenvalues of nonnormal Toeplitz
matrices tend to `trace out' strikingly
the pseudospectra of non-perturbed matrices. 
As suggested by Proposition \ref{thm:pseudoS},
the boundaries of the pseudospectra will be lined up
along the symbol curves of the corresponding
Toeplitz operators. 
Mathematical studies of  
\textit{banded} Toeplitz matrices with 
random perturbations have been reported 
\cite{BGKS22,BKMS21,BPZ19,BPZ20,BZ20,SV21}. 

We notice that since
$J$ in the present perturbation-term
$\delta J$ is the all-ones matrix, 
the Toeplitz operators
corresponding to $S^{(1)}_{\delta J}(m)$, $m=1,2,\dots,n$,
are \textit{not banded} and hence
symbols can not be defined. 
We consider, however, that the inner circle and annuli
found in the numerical results, 
Figs.~\ref{fig:model1_numerical_m2}--\ref{fig:model1_numerical_m15},
represent the boundaries and structures of the 
$\varepsilon$-pseudospectra with appropriate values 
of $\varepsilon$.

Figure~\ref{fig:ps} shows the 
contour plots of the 2-norms of
the resolvents, 
$\|(zI-S^{(1)}_{\delta J}(m))^{-1}\|$, 
for $m=1,2,3$, and 4, where $n=50$ and $\delta=0.01$.
Here the dots in the outer regions denote
the exact eigenvalues.
The values of $\|(zI-S^{(1)}_{\delta J}(m))^{-1}\|$
grow exponentially up to $10^5$
as we approach to the origin. 
With a given small value of $\varepsilon$
the $\varepsilon$-pseudospectrum
decreases monotonically
as $m$ increases.

\subsection{Model 2}
\label{sec:model2_pseudo}

\begin{figure*}
\begin{tabular}{cccc}
      \begin{minipage}[t]{0.23\hsize}
        \centering
        \includegraphics[keepaspectratio, scale=0.20]
        {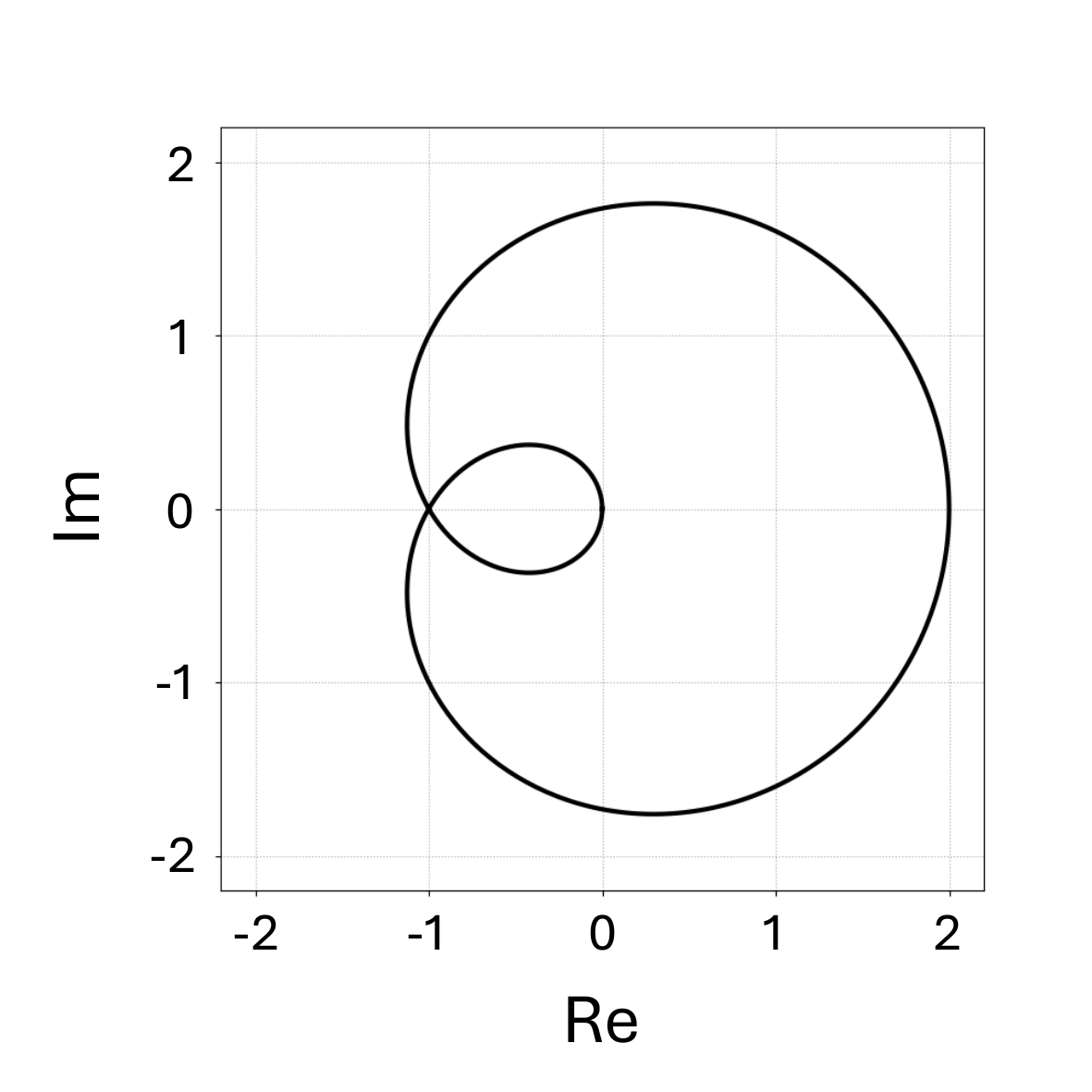}
        \subcaption{$m=1$}
        \label{fig:symbol2_m1}
      \end{minipage} &
      \begin{minipage}[t]{0.23\hsize}
        \centering
        \includegraphics[keepaspectratio, scale=0.20]
        {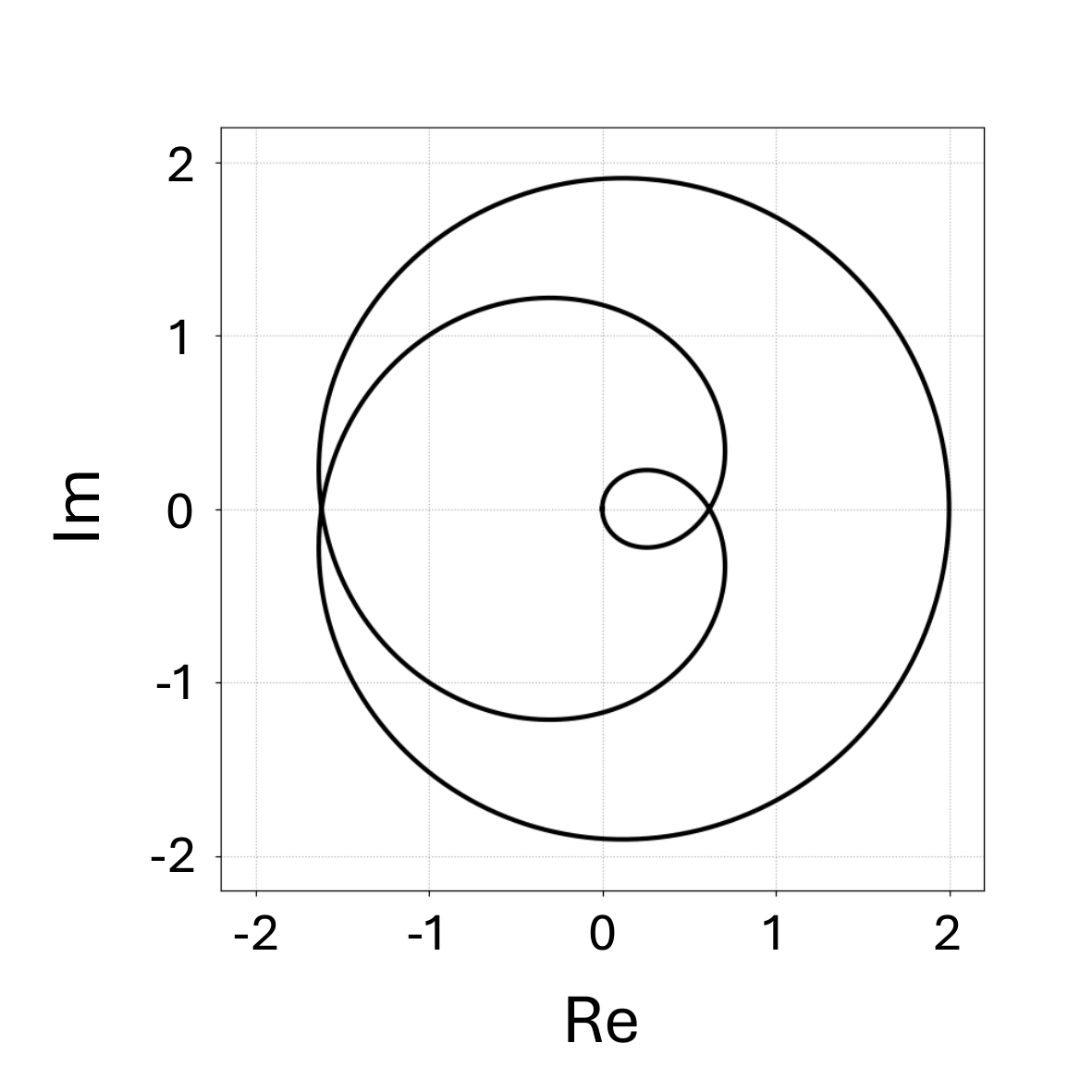}
        \subcaption{$m=2$}
        \label{fig:symbol2_m2}
      \end{minipage} &
      \begin{minipage}[t]{0.23\hsize}
        \centering
        \includegraphics[keepaspectratio, scale=0.20]
        {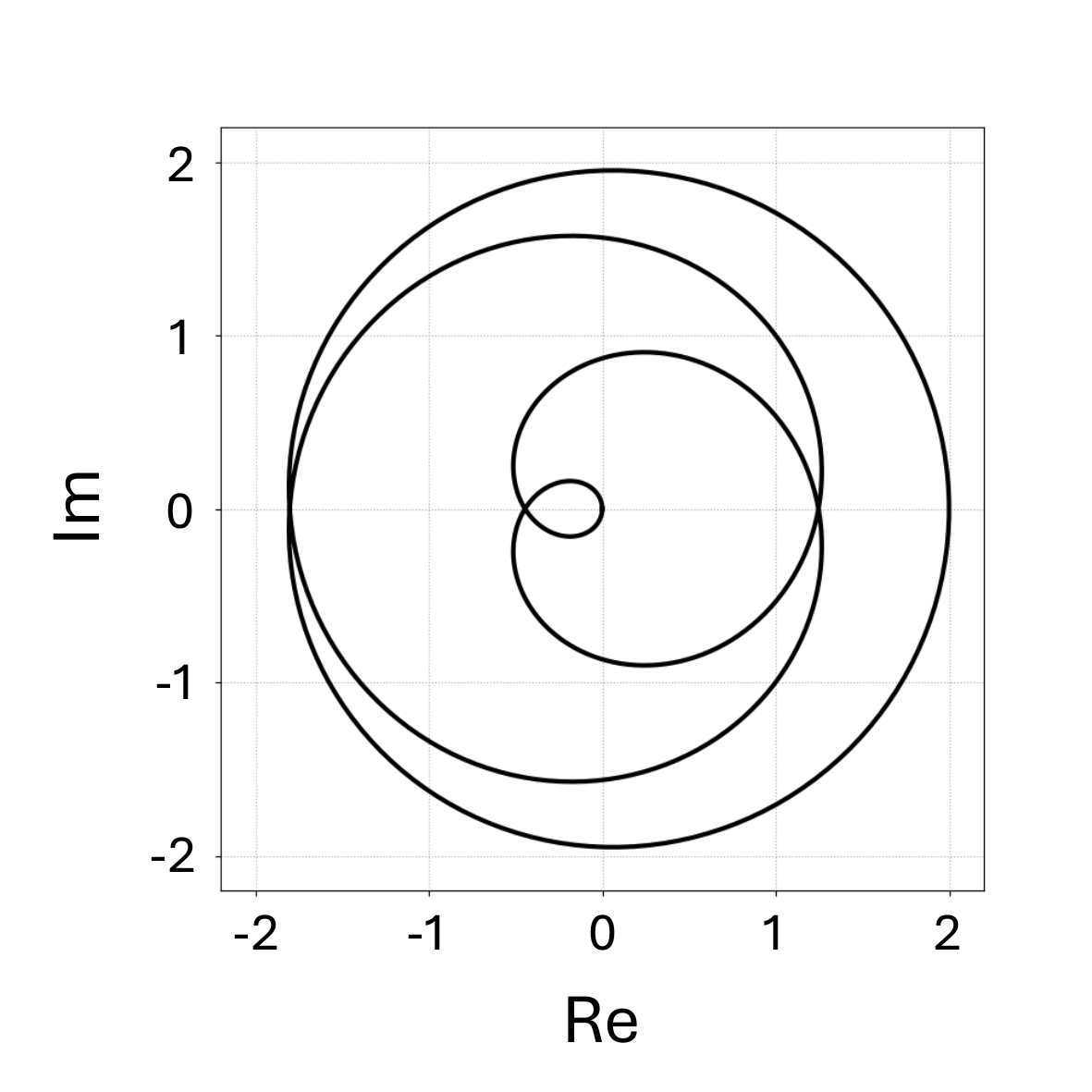}
        \subcaption{$m=3$}
        \label{fig:symbol2_m3}
      \end{minipage} &
      \begin{minipage}[t]{0.23\hsize}
        \centering
        \includegraphics[keepaspectratio, scale=0.20]
        {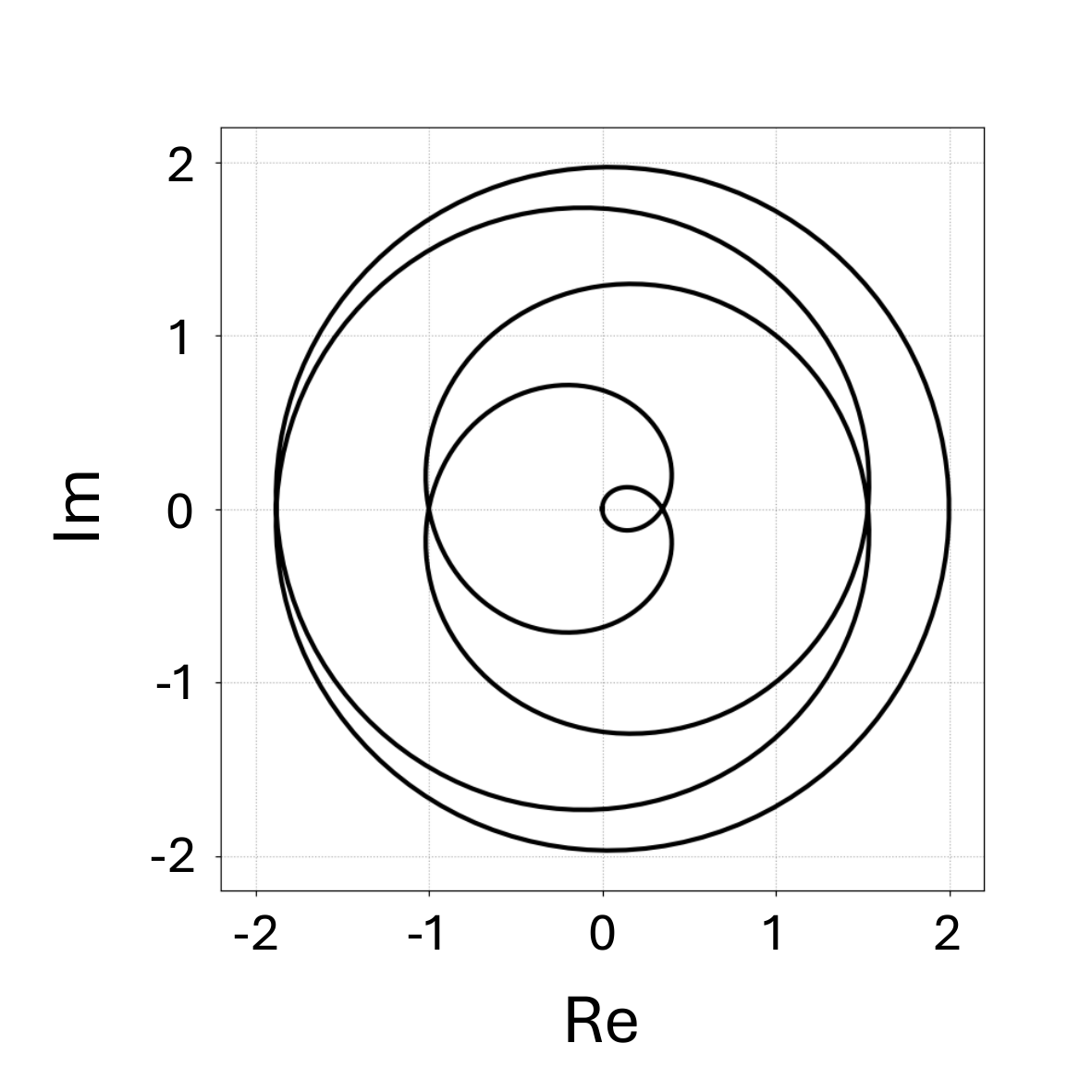}
        \subcaption{$m=4$}
        \label{fig:symbol2_m4}
      \end{minipage} 
    \end{tabular}
     \caption{The symbol curves of 
     $S^m+aS^{m+1}$ with $a=1$
     at $m=1,2,3$, and 4.}
\label{fig:symbols_model2}

\begin{tabular}{cccc}
      \begin{minipage}[t]{0.23\hsize}
        \centering
        \includegraphics[keepaspectratio, scale=0.18]
        {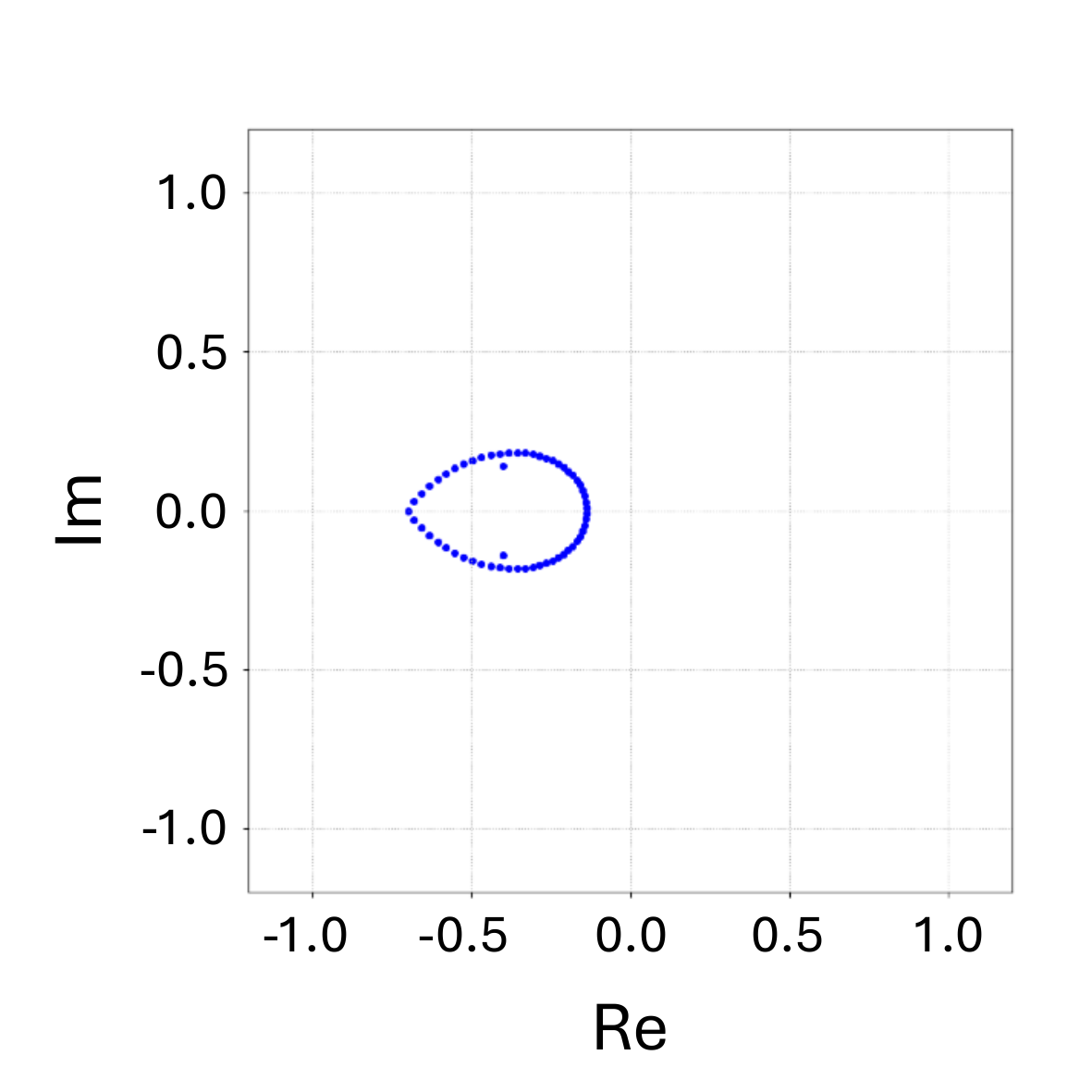}
        \subcaption{$m=1$}
        \label{fig:inner_m=1}
      \end{minipage} &
      \begin{minipage}[t]{0.23\hsize}
        \centering
        \includegraphics[keepaspectratio, scale=0.18]
        {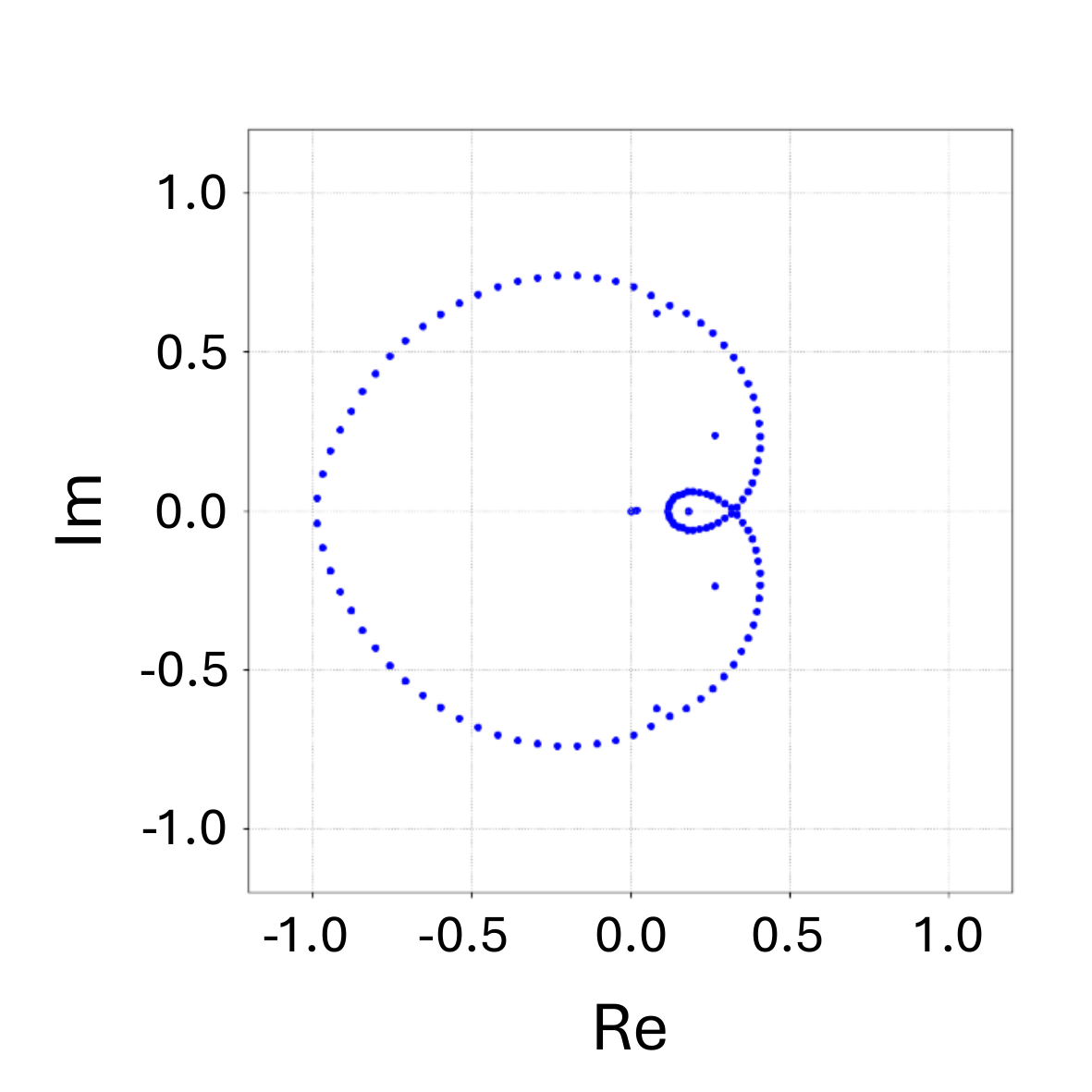}
        \subcaption{$m=2$}
        \label{fig:inner_m=2}
      \end{minipage} &
      \begin{minipage}[t]{0.23\hsize}
        \centering
        \includegraphics[keepaspectratio, scale=0.18]
        {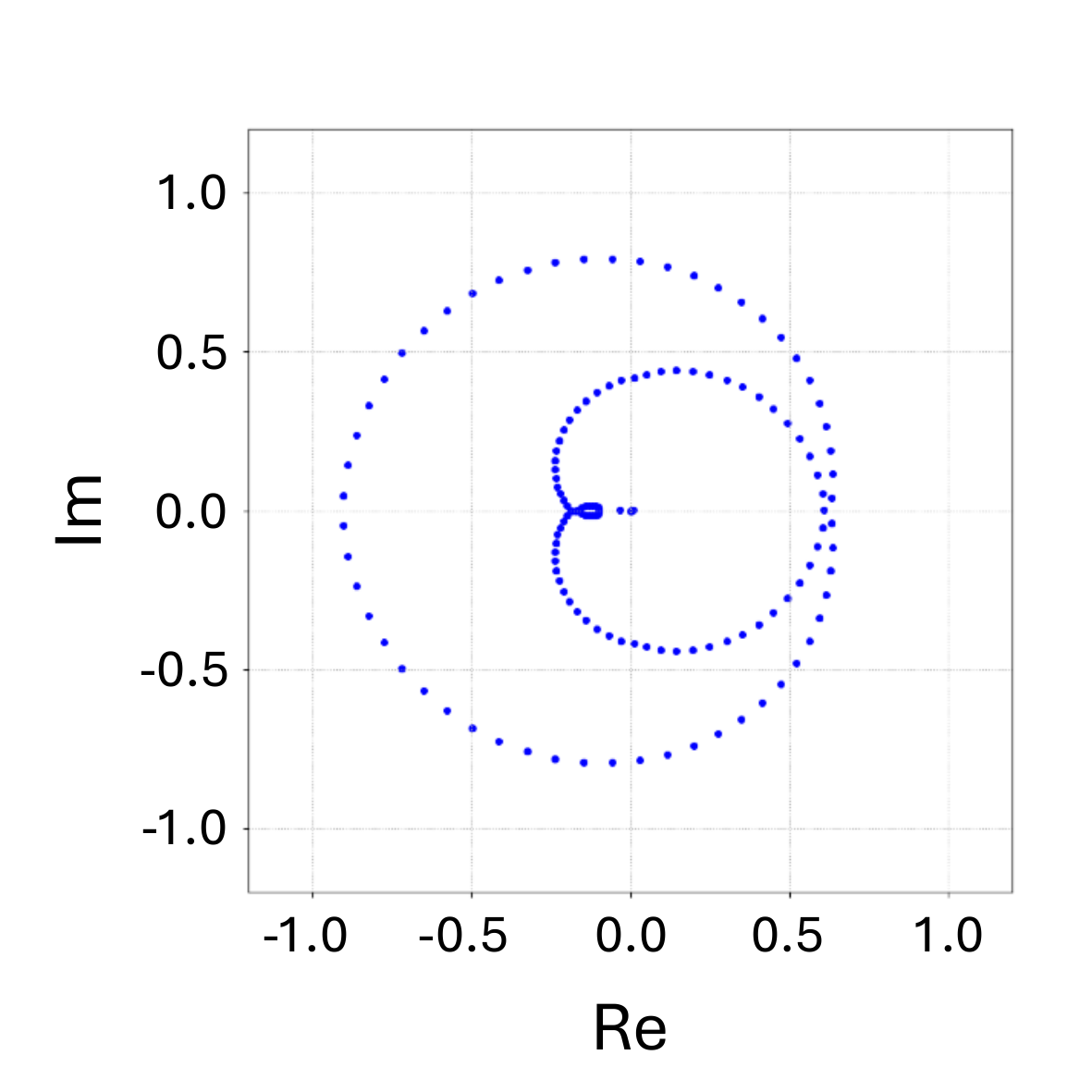}
        \subcaption{$m=3$}
        \label{fig:inner_m=3}
      \end{minipage} &
      \begin{minipage}[t]{0.23\hsize}
        \centering
        \includegraphics[keepaspectratio, scale=0.18]
        {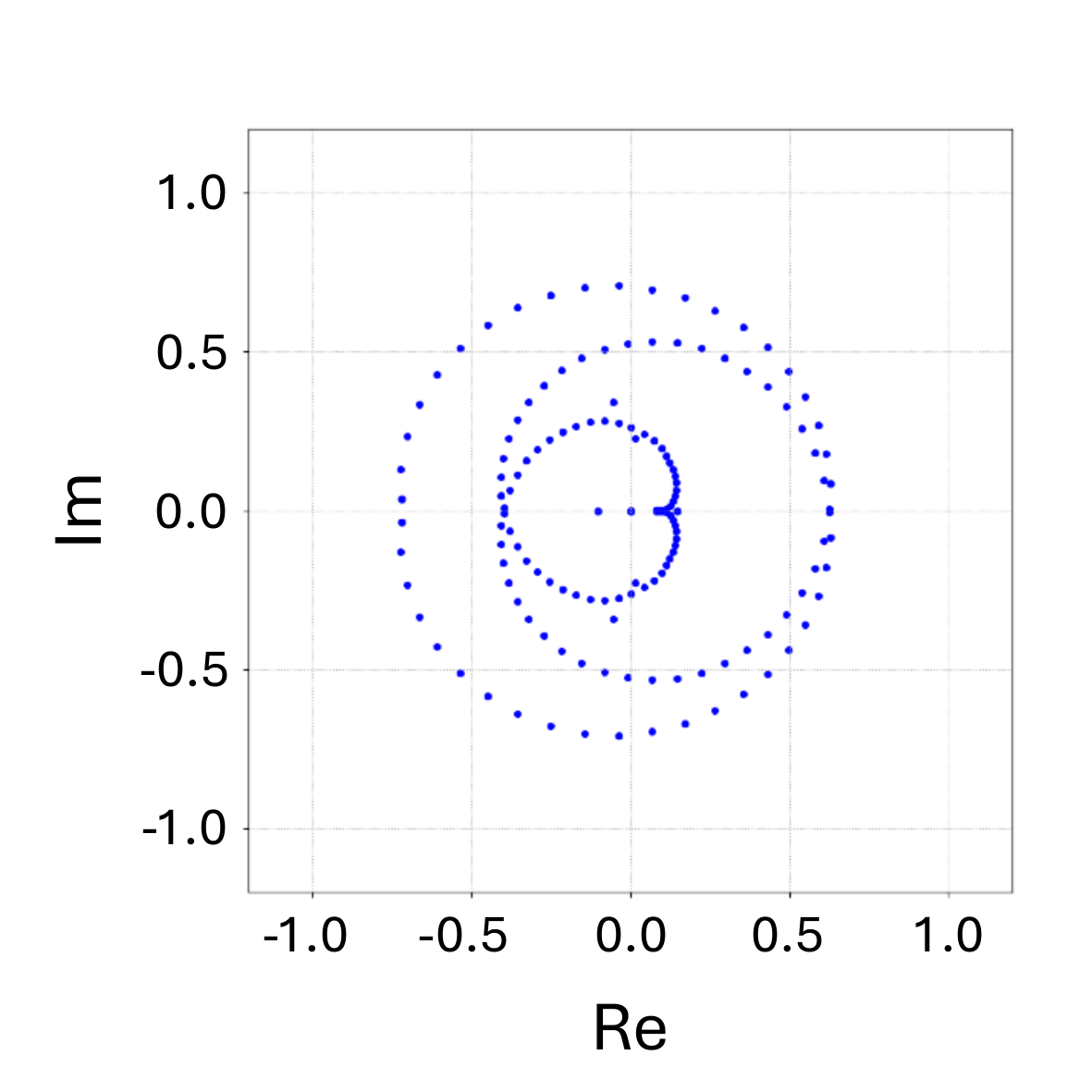}
        \subcaption{$m=4$}
        \label{fig:inner_m=4}
      \end{minipage} 
    \end{tabular}
     \caption{The inner parts of the numerically obtained
     eigenvalues for \textbf{model 2}
     showing structures of pseudospectra.}
\label{fig:inner_ps_model2}
\end{figure*}

Let $\widehat{S}^m+ a \widehat{S}^{m+1}$,
$m \in \N$ be the 
Toeplitz operators corresponding to
$S^m+a S^{m+1}$, $m \in \{1,2, \dots, n \}$.
The symbols are given by
\[
f_{\widehat{S}^m+a \widehat{S}^{m+1}}(z) 
=z^m+a z^{m+1},
\quad m \in \N.
\]
In Fig.~\ref{fig:symbols_model2},
we show the symbol curves
$f_{\widehat{S}^m+a \widehat{S}^{m+1}}(\T)$
with $a=1$ for $m=1, 2, 3 $, and 4.
We can consider that each symbol curve consists
of $m+1$ closed simple curves with different sizes.
Each closed simple curve is 
symmetric with respect to the real axis $\R$
and adjacent ones osculate at a point
on $\R$.
For instance, in the symbol curve 
for $m=3$ in Fig.~\ref{fig:symbol2_m3},
the osculating points of 
the composing closed simple curves are at about
$-1.8, 1.2$, and $-0.4$. 
We consider to separate them into two parts;
a single \textit{outermost closed simple curve} 
and an \textit{inner part} composed of
$m$ smaller closed simple curves osculating each other. 

Now we consider such separation also for the plots
of numerically obtained eigenvalues
shown in Fig.~\ref{fig:model2_numerical_ms}.
We have observed the following.

\begin{description}
\item{\rm (i)} \,
At each time $m$, 
the dots composing
the outermost curve
shown in the numerical result,
Fig.~\ref{fig:model2_numerical_ms}, 
coincide with the dots in
the outermost curve 
consisting of the exact eigenvalues
shown in Fig.~\ref{fig:model2_exact_ms}.

\item{\rm (ii)} \,
We cut out 
the inner parts composed by dots
in the numerical result, Fig.~\ref{fig:model2_numerical_ms}, 
and show them in Fig.~\ref{fig:inner_ps_model2}.
They seem to be the scale-downs of the
inner parts of the symbol curves 
$f_{\widehat{S}^m+a \widehat{S}^{m+1}}(\T)$
shown in Fig.~\ref{fig:symbols_model2}. 
Such structures can not be found
in Fig.~\ref{fig:model2_exact_ms} plotting the
exact eigenvalues. 
The size of the inner part decreases rapidly
as $m$ increases and the complicated patterns 
of the inner parts are smeared out for large values of $m$.

\item{\rm (iii)} \,
Many of the exact eigenvalues on $\R_-$ found in
Fig.~\ref{fig:model2_exact_ms} are hidden inside of
the inner parts.
Additional dots, which are not found in Fig.~\ref{fig:model2_exact_ms}, 
are observed only inside the inner parts. 

\end{description}

Based on the above observations, we give 
the following conjecture for \textbf{model 2}.

\begin{con}
\label{thm:conjecture1}
For \textbf{model 2}, the following holds.
\begin{description}
\item{\textup{(i)}} \,
At each time $m$, 
the dots composing
the outermost curve
shown in the numerical result,
Fig.~\ref{fig:model2_numerical_ms}, 
are exact eigenvalues of $S_{\delta J}^{(2)}(m)$.

\item{\textup{(ii)}} \,
The inner parts composed by dots
in the numerical result, Fig.~\ref{fig:model2_numerical_ms}, 
are not exact eigenvalues of $S_{\delta J}^{(2)}(m)$ for $m \geq 2$.
They are the eigenvalues of the perturbed system
of $S_{\delta J}^{(2)}(m)$ in which uncontrolled 
rounding errors of computer are added. 
The distributions of the eigenvalues
of such perturbed system represent 
structures of pseudospectrum including
$\lambda_0$ of the original system. 
They reflect the inner parts of the spectra
of the Toeplitz operator, 
$\sigma(\widehat{S}^m+a\widehat{S}^{m-1})$, 
without the deterministic perturbation by
$\delta J$ nor uncontrolled perturbations
by rounding errors of computer. 
The size of the pseudospectrum 
including $\lambda_0$, which is
represented by the inner part of symbol curve, 
decreases exponentially as $m$ increases.
\end{description}
\end{con}

\SSC{Asymptotics in Infinite Matrix Limits}
\label{sec:asymptotics}

For \textbf{model 1}, 
we fix $m$ and $\delta$ so that they
satisfy $m \leq n-1$ and $\delta > 4m/n^2$. 
Proposition \ref{thm:ev2} (ii)
implies 
\begin{align}
\lambda_1(m) &\sim
\frac{n \delta}{2} + 1 +
\frac{n \delta}{2} 
\sqrt{1-\frac{4m}{n^2 \delta} }
\nonumber\\
&= n \delta +1 -\frac{m}{n}
+\rO(n^{-2}) \to \infty, 
\label{eq:large_ev2}
\end{align}
as $n \to \infty$.
Here we have used the 
formula of the generating function for
the Catalan numbers \cite{Rio68}, 
$\sum_{k=0}^{\infty} C_k x^k
=(1-\sqrt{1-4 x})/(2x)$ for
$x < 1/4$. 
The fact
\eqref{eq:large_ev2} implies that $\lambda_1(m)$ 
solves the quadratic equation asymptotically, 
\begin{equation}
z^2-(n \delta +2) z + 1 + 
n \delta (1+ m/n)=0, 
\label{eq:asympto1}
\end{equation}
as $n \to \infty$.
We write the solution of \eqref{eq:asympto1}
other than \eqref{eq:large_ev2} as 
$\widetilde{\lambda}_1(m)$,
which behaves as
\begin{align}
\widetilde{\lambda}_1(m) &=
(n \delta)/2+1- (n \delta/2)
\sqrt{1- 4m/(n^2 \delta)}
\nonumber\\
&= 1+ m/n +\rO(n^{-2}) \to 1, 
\label{eq:large_ev2B}
\end{align}
as $n \to \infty$. 
Then it is easy to verify that 
the left-hand-side of
Eq.~\eqref{eq:eigenvalues2} in Theorem \ref{thm:ev1}
is written as
\begin{align}
&(z-\lambda_1(m)) (z-\widetilde{\lambda}_1(m))
\sum_{k=0}^{p_1-1} (p_1-k) z^k 
\nonumber\\
& \quad +(p_1+1)
[ z - \{ n \delta +1 - 1/(p_1+1) \} ].
\label{eq:factorize}
\end{align}
We notice that
\begin{align*}
&\{ (p_1+1)/p_1 \}
[ z - \{ n \delta + 1 - 1/(p_1+1) \} ]
\nonumber\\
& \quad = z-(n \delta +1) - n \delta/p_1 
+\rO(n^{-1})
\nonumber\\
& \quad = \left\{ z - \left( n \delta + 1 - m/n 
+\rO(n^{-2}) \right) \right\} 
\nonumber\\
& \qquad \times
( 1 + 1/p_1 + \rO(n^{-2}))
\nonumber\\
& \quad \sim 
(z-\lambda_1(m)) ( 1 + 1/p_1), 
\end{align*}
as $n \to \infty \, \, (p_1 \to \infty)$
for \eqref{eq:large_ev2}.
Then \eqref{eq:factorize} is factorized by $z-\lambda_1(m)$
asymptotically in the sense that
\begin{align}
&\frac{1}{p_1} \left[
z^{p_1+1}-n \delta \sum_{k=0}^{p_1}
\{ 1- (p_1-k) m/n\} z^k \right]
\nonumber\\
& \,
\sim (z-\lambda_1(m))
\Big\{
(z-\widetilde{\lambda}_1(m)) 
\nonumber\\
& \qquad \times
\sum_{k=0}^{p_1-1}
(1- k/p_1) z^k+1+1/p_1 \Big\}, 
\label{eq:model1_asym}
\end{align}
as $n \to \infty \, (p_1 \to \infty)$.

\begin{prop}
\label{thm:model1_asym}
For \textbf{model 1}, 
fix $m$ and $\delta$ satisfying
$m \leq n-1$ and $\delta > 4m/n^2$.
Then as $n \to \infty$, $p_1$ non-zero exact eigenvalues 
except the outlier $\lambda_1(m)$ 
tend to be well approximated by
\[
e^{2 \pi i \ell/(p_1+1)},
\quad \ell=1,2, \dots, p_1.
\]
That is, the eigenvalues 
become to form a configuration such that
one point at $z=1$ is eliminated from the 
equidistance $p_1+1$ points
$\{e^{2 \pi i \ell/(p_1+1)}; \ell=0,1, \dots, p_1\}$
on $\T$.
\end{prop}
\vskip 0.3cm
\noindent{\it Proof} \,
For \eqref{eq:large_ev2B} 
$\widetilde{\lambda}_1(m) \in \R$ and
$\widetilde{\lambda}_1(m) \to 1$ as $n \to \infty$.
Eq.~\eqref{eq:model1_asym} 
including the term $1+1/p_1$ in the right-hand-side 
implies that
$\widetilde{\lambda}_1(m)$ does \textit{not} satisfy 
Eq.~\eqref{eq:eigenvalues2} in Theorem \ref{thm:ev1}.
By \eqref{eq:large_ev2B} and the
summation formulas \eqref{eq:expansion}, we see that
\begin{align*}
&(z-\widetilde{\lambda}_1(m)) \sum_{k=0}^{p_1-1} 
\left( 1 - \frac{k}{p_1} \right) z^k+1+\frac{1}{p_1}
\nonumber\\
& \, 
\sim \left( z-1-\frac{m}{n} \right) \left\{ \frac{1}{1-z}
-\frac{z(1-z^{p_1})}{p_1 (1-z)^2} \right\}+1+\frac{1}{p_1}
\nonumber\\
& \, 
\sim -\frac{m}{n} \frac{1}{1-z}
+ \frac{z(1-z^{p_1})}{p_1(1-z)}+\frac{1}{p_1}
\nonumber\\
& \, 
=-\frac{1}{p_1 (1-z)}
\left[ z^{p_1+1} - \left( 1 - \frac{m}{n} p_1 \right) \right], 
\end{align*}
as $n \to \infty \, \, (p_1 \to \infty)$.
We write the solution of the equation
\[
z^{p_1+1}= 1 - \frac{m}{n} p_1
\]
as $z=r e^{i \theta}$, $r >0$, $\theta \in [0, 2 \pi)$.
Then we have
\begin{align*}
&\log r = \frac{1}{p_1+1} \log \left(
1- \frac{m}{n} p_1 \right),
\nonumber\\
&(p_1+1) \theta = 0 \quad
\mbox{mod $2 \pi$}.
\end{align*}
They give $r \to 1$ as $n \to \infty$ ($p_1 \to \infty$)
and $\theta=2 \pi \ell/(p_1+1)$, 
$\ell=1,2, \dots, p_1$.
Here the case $\theta=0$ ($\ell=0$) should not be
included,
since $\widetilde{\lambda}_1(m)$ is not
the solution as mentioned above.
The assertion is hence proved. 
\qed
\vskip 0.3cm

Now we consider the quadratic equation for $z$, 
\begin{align}
&\left( \frac{z}{1+a} \right)^2
-\left( \frac{n \delta}{1+a} +2 \right)
\frac{z}{1+a}
\nonumber\\
& \quad 
+1 + \frac{n \delta}{1+a}
\left\{ 1 + \frac{m}{n}+\frac{a}{(1+a)n} \right\}
=0,
\label{eq:Eq1}
\end{align}
which is obtained from \eqref{eq:asympto1} by the
replacement \eqref{eq:replace1}. 
We write the solutions of \eqref{eq:Eq1} as
\[
\kappa_{\pm}
:= \frac{n \delta}{2}+1+a
\pm \frac{n \delta}{2}
\sqrt{1- \frac{4\{ (1+a) m +a \}}{n^2 \delta}}.
\]
Notice that
\begin{align*}
&\kappa_{+} = n \delta + 1 + a 
\nonumber\\
& \quad -
(1+a) \sum_{k=0}^{\infty} C_k 
\left( 
\frac{m}{n}+\frac{a}{(1+a)n}
\right)^{k+1} 
\left( \frac{1+a}{n \delta} \right)^k,
\end{align*}
where $C_k$, $k \in \N_0$, are the Catalan numbers 
\eqref{eq:Catalan}.
The first three lines of \eqref{eq:large_evG} 
in Proposition \ref{thm:evG2} (ii) for $\lambda_1(m)$
of \textbf{model 2} are regarded as a truncation 
of the infinite series of $\kappa_+$.
Then the first two lines of the left-hand-side 
of \eqref{eq:eigenvaluesG2} in Theorem \ref{thm:evG1}
are written as follows,
\begin{align*}
& \left( \frac{z}{1+a} \right)^{p_1+1}
-\frac{n \delta}{1+a} \sum_{k=0}^{p_1}
\nonumber\\
& \, 
\times
\left[ 1 - (p_1-k) 
\left\{ \frac{m}{n} + \frac{a}{(1+a)n} \right\}
\right]
\left( \frac{z}{1+a} \right)^{k}
\nonumber\\
& \, 
=\left( \frac{z}{1+a} -\frac{\kappa_+}{1+a} \right)
\left( \frac{z}{1+a} -\frac{\kappa_-}{1+a} \right)
\nonumber\\
& \quad \times
\sum_{k=0}^{p_1-1} (p_1-k)
\left( \frac{z}{1+a} \right)^k
\nonumber\\
& \quad
+(p_1+1) \left\{
\frac{z}{1+a} - \left(
\frac{n\delta}{1+a}+1-\frac{1}{p_1+1} \right) \right\}.
\end{align*}
This equality can be regarded as the extension
of \eqref{eq:factorize} including $a$
obtained by the replacement \eqref{eq:replace1}. 

Lemma \ref{thm:p_0_1} clarified the condition for $m$
so that $p_1=p_2$; that is,
$m \in I_{n-1} \setminus T_{n-1}$.
Since $\min I_{n-1}=\lceil \sqrt{n-1} \, \rceil$, 
if $m \gtrsim \min I_{n-1}$, then 
$p_1=[(n-1)/m] \to \infty$ as $n \to \infty$.
Proposition \ref{thm:model1_asym}
is generalized for \textbf{model 2}
in such a situation. 
\begin{prop}
\label{thm:model2_asym}
For \textbf{model 2}, 
fix $\delta$ and $a$. Consider $m$ satisfying
$\delta > 4\{ (1+a) m + a \}/n^2$ 
and $p_1 =p_2$. 
Then as $n \to \infty$, $p_1$ non-zero exact eigenvalues 
except the outlier $\lambda_1(m)$ 
become to be well approximated by
\[
(1+a) e^{2 \pi i \ell/(p_1+1)},
\quad \ell=1,2, \dots, p_1.
\]
That is, the eigenvalues 
become to form a configuration such that
one point at $z=1+a$ is eliminated from the 
equidistance $p_1+1$ points
$\{(1+a) e^{2 \pi i \ell/(p_1+1)}; \ell=0,1, \dots, p_1\}$
on $\T_{1+a}$. 
\end{prop}

We have performed numerical calculations
of \textbf{model 1} and \textbf{model 2} 
for a variety of the matrix size $n$
with $m \in \{1,2, \dots, n\}$.
Then we arrived at the following conjectures.

\begin{con}
\label{thm:conjecture2}
For \textbf{model 1}, 
at each fixed $m \geq 1$, 
the boundary of the pseudospectrum
including $\lambda_0$
increases its size as $n$ increases. 
It converges to the unit circle 
$\T=f_{\widehat{S}^m}(\T)$ 
as $n \to \infty$.
The inside of $\T$ becomes to be fulfilled
by the eigenvalues of perturbed system 
as $n \to \infty$.
\end{con}

\begin{con}
\label{thm:conjecture3}
For \textbf{model 2}, the following holds. 
\begin{description}
\item{\textup{(i)}} \,
At each fixed $m \geq 1$, 
the exact eigenvalues composing outermost curve
converge to the outermost closed simple curve 
of the symbol curve 
$f_{\widehat{S}^m+a \widehat{S}^{m+1}}(\T)$
as $n \to \infty$. These eigenvalues are insensitive
and robust to random perturbations and rounding errors 
of computer.

\item{\textup{(ii)}} \,
At each fixed $m \geq 1$, 
the distribution of eigenvalues of
perturbed system observed in the inner part 
represents the pseudospectrum
including $\lambda_0$ of $S_{\delta J}^{(2)}(m)$.
The size of the pseudospectrum
including $\lambda_0$ increases as $n$ increases.
It converges to the 
inner part of the the symbol curve 
$f_{\widehat{S}^m+a \widehat{S}^{m+1}}(\T)$
as $n \to \infty$.
Only inside of the inner part of the the symbol curve 
$f_{\widehat{S}^m+a \widehat{S}^{m+1}}(\T)$
becomes to be fulfilled
by the eigenvalues of perturbed systems 
as $n \to \infty$.
\end{description}
\end{con}

Remark 4 and Lemma \ref{thm:p_0_1} in
Section \ref{sec:model2_ev} suggest that 
in the period 
$1 \leq m < \lceil \sqrt{n-1} \, \rceil$, 
$p_1 \geq p_2+1$ and the last part of the left-hand-side
of \eqref{eq:eigenvaluesG1} should play an important
role to determine the exact eigenvalues of \textbf{model 2}.
For such $m$,
however, the ratio $m/n$ becomes zero
as $n \to \infty$. 
Hence, when we consider the case in which
both of $n$ and $m$ are sufficiently large
with a fixed value of the ratio $m/n >0$,
the last part of the left-hand-side
of \eqref{eq:eigenvaluesG1} becomes irrelevant 
to determine the exact eigenvalues.
Lemma \ref{thm:p_0_1} implies that if
$m \in I_{n-1} \cap T_{n-1}$, 
then $p_1-p_2=1$ and the last part of the left-hand-side
of \eqref{eq:eigenvaluesG1} gives a term
which does not include $z$ but depends
on the value of $m$ and the parameter $a$.
It is obvious that for each 
$m =[(n-1)/k] \in I_{n-1} \cap T_{n-1}$, $k=1,2,\dots$, 
we have $m/n \to 1/k$ as $n \to \infty$. 
We also notice that the additional term
$a/\{(1+a)n\}$ for the last formula in the
replacement \eqref{eq:replace1} becomes zero
as $n \to \infty$ with fixed $a$.
From the numerical calculations and
the above considerations, 
we have the following conjecture.

\begin{con}
\label{thm:conjecture4}
For both of \textbf{model 1} and \textbf{model 2}, 
the sizes of the spectra and the pseudospectra are determined by
the ratio $m/n$, if both of $n$ and $m (\leq n)$ are 
sufficiently large and the other parameters
$\delta$ and $a$ are fixed.
Hence in the numerically observed eigenvalues with
different values of the pair $(n, m)$ but with
the same ratio $n/m$, 
we will observe the patterns with the components
which are in the similar sizes with each other, 
if $n$ and $m (\leq n)$ are both sufficiently large. 
In other words, 
we will have nontrivial scaling limits
$n \to \infty$ and $m \to \infty$
with fixed $m/n \in (0,1)$.
\end{con}
This conjecture is partially proved by calculating
the resolvent of $S_{\delta J}^{(2)}(m)$
in \cite{MKS24b}.

\SSC{Concluding Remarks and Future Problems}
\label{sec:future}

\begin{figure}[htbp]
    \begin{tabular}{cc}
      \begin{minipage}[t]{0.45\hsize}
        \centering
        \includegraphics[keepaspectratio, scale=0.12]
        {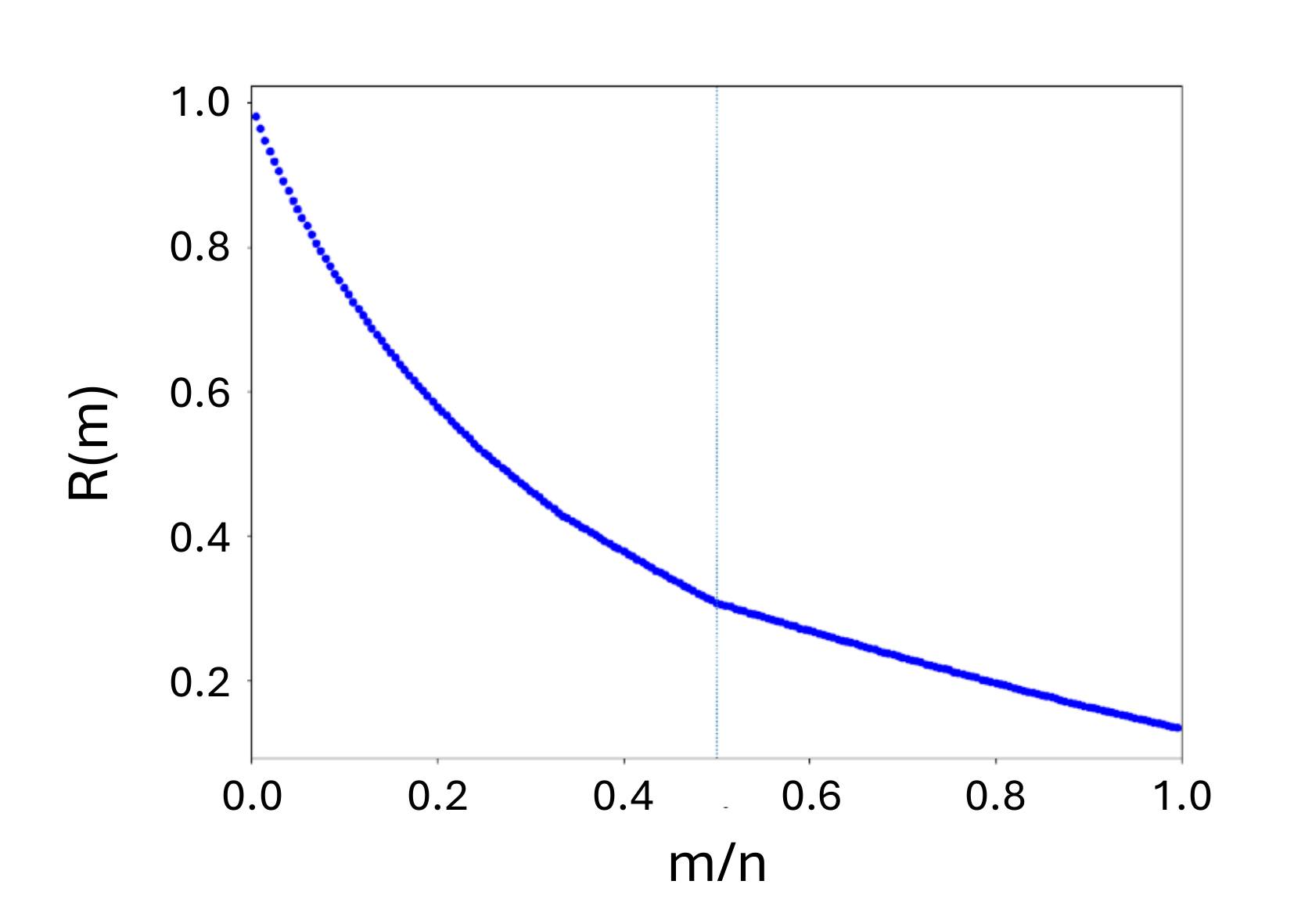}
        \subcaption{}
        \label{fig:Gauss1}
      \end{minipage} 
 
      \begin{minipage}[t]{0.45\hsize}
        \centering
        \includegraphics[keepaspectratio, scale=0.12]
        {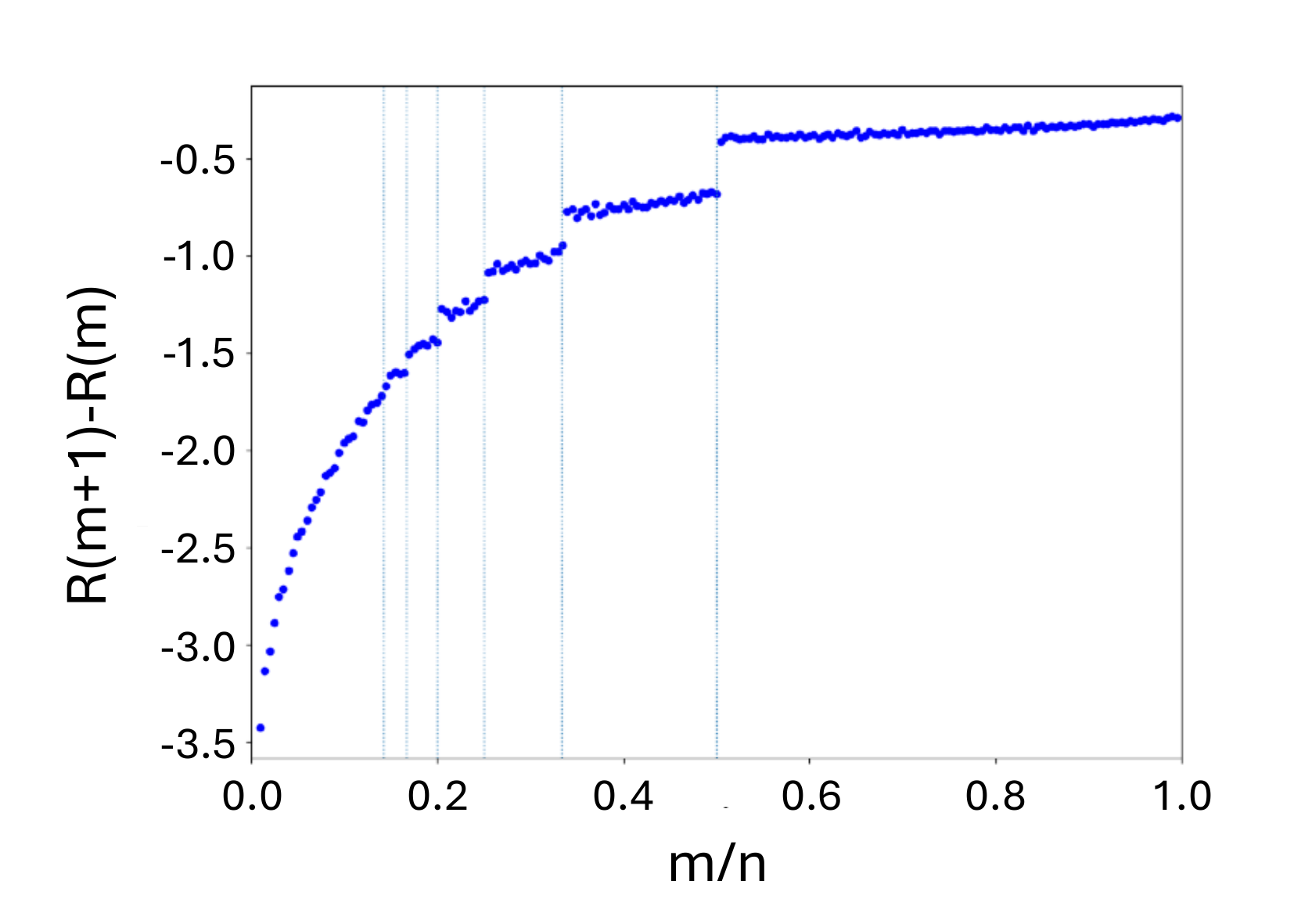}
        \subcaption{}
        \label{fig:Gauss_Devil}
      \end{minipage} 
\\
       \begin{minipage}[t]{0.45\hsize}
        \centering
        \includegraphics[keepaspectratio, scale=0.12]
        {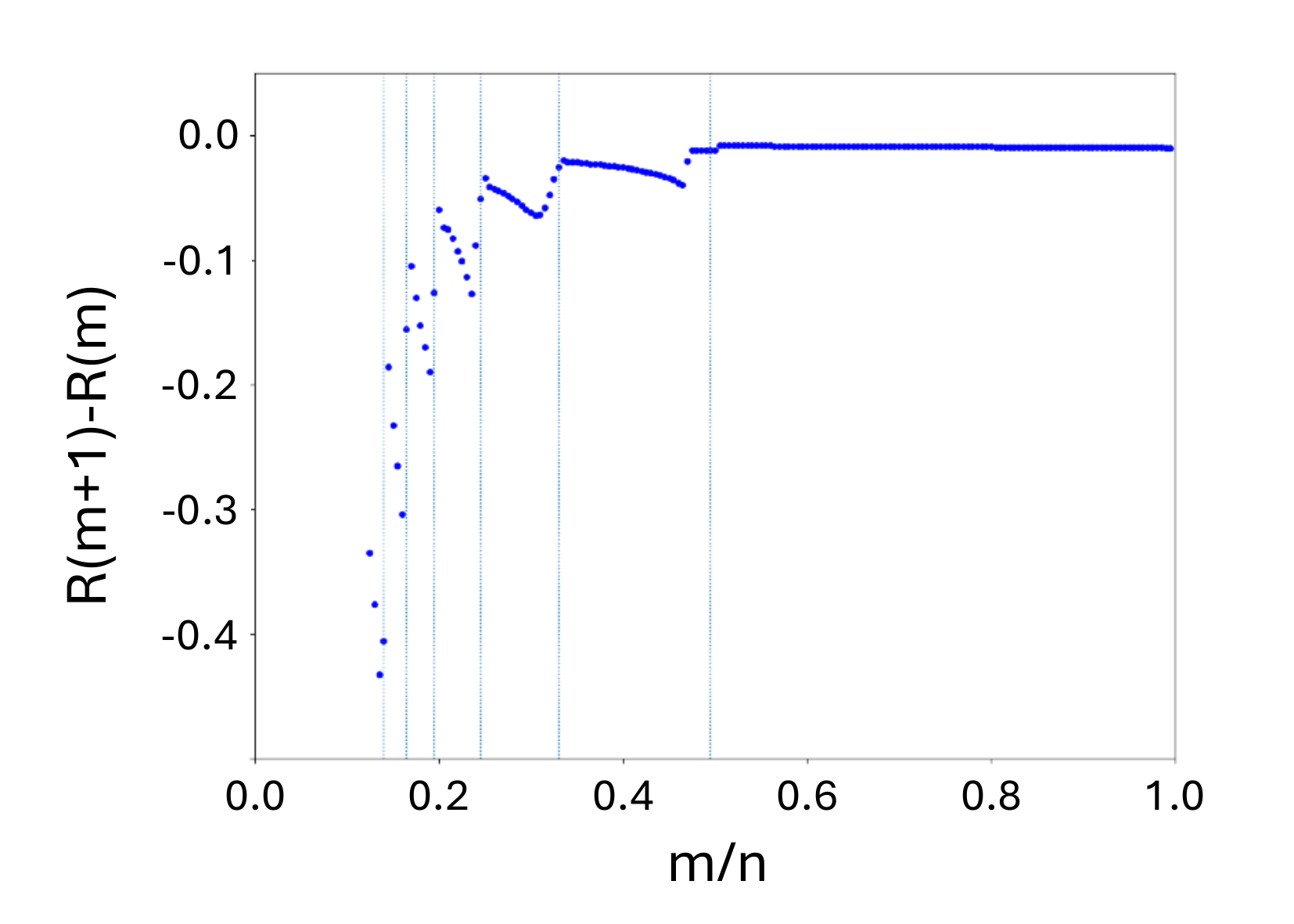}
        \subcaption{}
        \label{fig:model1_Devil}
      \end{minipage} 

        \begin{minipage}[t]{0.45\hsize}
        \centering
        \includegraphics[keepaspectratio, scale=0.12]
        {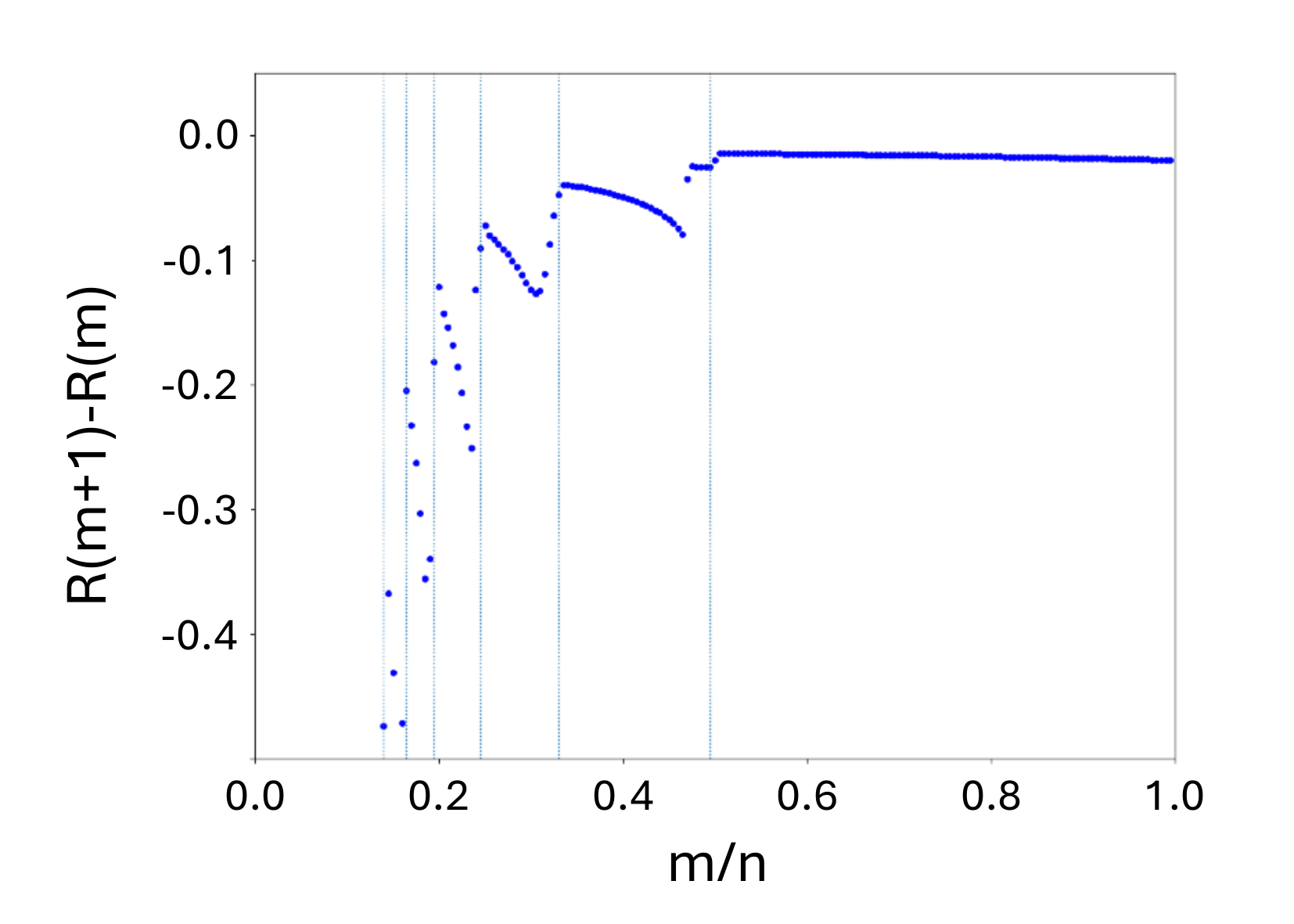}
        \subcaption{}
        \label{fig:model2_Devil}
      \end{minipage} 
\end{tabular}
     \caption{
     (a) Dependence on the ratio $m/n$ of the mean radius $R(m)$
        of eigenvalues for the random process
        $(S_{\delta Z}(m))_{m=1}^n$ for $n=200$ and
        $\delta=0.01$ evaluated by averaging over
        $10^3$ sample processes.
     (b) Devil's staircase-like structure
     observed in the first derivative
     $R(m+1)-R(m)$ for the random process
     $(S_{\delta Z}(m))_{m=1}^n$.
     (c) Deformed version of staircase structure 
     observed in the   
     first derivative $R(m+1)-R(m)$ for \textbf{model 1}.
     (d) Deformed version of staircase structure 
     observed in the   
     first derivative $R(m+1)-R(m)$ for \textbf{model 2}.
      }
\label{fig:Gauss}
  \end{figure}

Now we discuss a possibility
to study the discrete-time \textit{random} process
$(S_{\delta Z}(m))_{m=1}^n$ defined as
\eqref{eq:pZ} by comparing it with the present models.
When we numerically simulate the eigenvalue process
of $(S_{\delta Z}(m))_{m=1}^n$,
we observe that the obtained dots form an annulus
which seems to be similar to Fig.~\ref{fig:annulus}.

The size of annulus, however, rapidly decreases as 
$m$ increases and 
the annulus becomes a shrinking disk.
We have calculated the \textit{mean radius} 
of annulus or disk at each time $m$,
which is defined as the mean of radial coordinates
of all dots of numerically obtained eigenvalues.
We have performed $10^3$ independent simulations
and averaged the time-dependence of mean radius.
As shown by Fig.~\ref{fig:Gauss1}, 
the expectation of mean radius, $R(m)$, 
which is evaluated by averaging over $10^3$
sample processes,
decreases monotonically in time $m$. 
Since a cusp is observed at $m/n=1/2$, we have calculated
the first derivative, that is, the increment
$R(m+1)-R(m)$, $m=1,2,\dots, n$, numerically.
The result shows 
\textit{devil's staircase-like} structure as shown in
Fig.~\ref{fig:Gauss_Devil},
where the thin vertical lines are given at
$m/n=1/k$, $k=2,3, \dots, 7$ from the right to the left.

As shown by Fig.~\ref{fig:model1_numerical_ms}
and Fig.~\ref{fig:ps} for \textbf{model 1}
and by Fig.~\ref{fig:model2_numerical_ms}
for \textbf{model 2}, 
we have observed the similar monotonic reduction of sizes
of exact-spectra and pseudospectra in time $m$.
As asserted by Conjecture \ref{thm:conjecture4},
this phenomenon will be described by the ratio $m/n$,
when $n$ and $m (\leq n)$ are both sufficiently large.
Figs.~\ref{fig:model1_Devil} and \ref{fig:model2_Devil}
show the $m/n$-dependence
of the fist derivatives of the mean radii of
all numerically observed eigenvalues
for \textbf{model 1} and \textbf{model 2}, respectively. 
We see deformed but similar staircase structures. 
In these figures, 
the thin vertical lines
are given at $m/n$ with 
times $m=[(n-1)/k]$, $k=1,2, \dots, 7$, 
which are included in $I_{n-1} \cap T_{n-1}$ and 
at which $p_1-p_2=1$ as proved by
Lemma \ref{thm:p_0_1} in Section \ref{sec:model2_ev}.
For $m \in I_{n-1} \cap T_{n-1}$, 
when $m \to m+1$
the degree $p_1+1$ of the polynomial equation
for the exact eigenvalues 
\eqref{eq:eigenvalues2} of \textbf{model 1}
(resp. \eqref{eq:eigenvaluesG2} of \textbf{model 2})
decreases by one, 
and at time $m$
the last part of the left-hand-side of
\eqref{eq:eigenvaluesG2} can give an
additional constant term to the polynomial equation
for \textbf{model 2}.
(For instance, when $n=200$, $p_1-p_2=1$
if $m=[(n-1)/k]=[199/k]$,
$k=1,2, \dots, 13$, which we conclude from
Lemma \ref{thm:p_0_1}.
The last part of the left-hand-side of
\eqref{eq:eigenvaluesG2} does not give any contribution,
however, 
when $k=1,2,4,5,8$, and 10, since the condition
$p_1 \geq (n+1)/(m+1)$ is not satisfied.)

The similarity of Fig.~\ref{fig:Gauss_Devil}
to Figs.~\ref{fig:model1_Devil} and \ref{fig:model2_Devil} 
suggests some connection between
the present deterministic processes
of non-banded Toeplitz matrices
with perturbations by rounding errors of computer
and the banded Toeplitz matrices
with random perturbations 
\cite{BGKS22,BKMS21,BPZ19,BPZ20,BZ20,SV21}.

We list out other two future problems.
\begin{description}
\item{(i)} \,
We have distinguished the exact non-zero eigenvalues
which are insensitive to perturbations 
from the numerically observed eigenvalues 
which are not the eigenvalues of the original model
$S_{\delta J}^{(\ell)}(m)$, 
but are the eigenvalues of perturbed systems
due to rounding errors of computer. 
The distributions of the latter eigenvalues
visualize the pseudospectrum including
$\lambda_0$ of the original model.
Mathematical proofs for Conjectures 
\ref{thm:conjecture1}, \ref{thm:conjecture2},
\ref{thm:conjecture3}, and \ref{thm:conjecture4}
will be challenging future problems. 
See \cite{MKS24b} for further considerations.
It will be an important subject in numerical analysis
to classify the eigenvalues systematically into two groups,
one of which is sensitive and unstable, and other one is
insensitive and robust with respect to 
perturbations \cite{CR01,HP05,TE05,WMG08}. 

\item{(ii)} \,
Recently non-Hermitian quantum physics has been
extensively studied \cite{AGU20}. 
We expect that the present mathematical study
of eigenvalue and pseudospectrum processes 
will be related to the \textit{pseudospectrum approaches}
to non-Hermitian quantum systems 
used in physics \cite{GAKTHU18,OKSS20}. 
\end{description}

\begin{footnotesize}
\vskip 0.3cm
\noindent{\bf Acknowledgements} \,
This study was carried out under the 
Open-Type Professional Development Program 
of the Institute of Statistical Mathematics, Tokyo
(2023-ISMHRD-7010) organized by
Takaaki Shimura, Satoshi Kuriki, 
Beno\^it Collins, Noriyoshi Sakuma, 
and Yuki Ueda.
The present authors would like to thank
J. Garza-Vargas for useful comments on the present work.
Part of the present study was done during the stay of the authors
at Institute for Mathematical Sciences,
National University of Singapore.
The authors thank Akira Sakai and Rongfeng Sun
for organizing the three-week fruitful program 
in December 2023.
This work was also supported by the Research Institute for Mathematical Sciences,
an International Joint Usage/Research Center located in Kyoto University.
MS was supported by JSPS KAKENHI Grant Numbers 
JP19K03674, 
JP21H04432,
JP22H05105,
JP23K25774, 
and
JP24K06888.
TS was supported by JSPS KAKENHI Grant Numbers 
JP20K20884,
JP21H04432, 
JP22H05105,
JP23K25774, 
and 
JP24KK0060.


\end{footnotesize}
\end{document}